\documentclass[nofootinbib]{revtex4}
\usepackage{amsmath}
\usepackage{amsfonts}
\usepackage{amssymb}
\usepackage{graphicx}
\usepackage{epstopdf}%
\newcommand{\Dslash}[1] { \setbox0=\hbox{$#1$}     
\dimen0=\wd0   \setbox1=\hbox{/} \dimen1=\wd1  \ifdim\dimen0>\dimen1        
 \rlap{\hbox to \dimen0{\hfil/\hfil}}  #1 \else \rlap{\hbox to \dimen1{\hfil$#1$\hfil}}  /  \fi  }

\newcommand{\bea}{\begin{eqnarray}}
\newcommand{\eea}{\end{eqnarray}}
\newcommand{\be}{\begin{equation}}
\newcommand{\ee}{\end{equation}}
\newcommand{\bs}{\boldsymbol}
\newcommand{\ns}{\Dslash{n}}
\newcommand {\nbs}{\Dslash{\bar n}}
\newcommand{\nbn}{\frac{\nbs\ns}{4}}

\begin{document}

\noindent
MKPH-T-12-09\\
HIM-2012-01
\bigskip

\title{
Factorizing  the hard and soft spectator scattering contributions  for the nucleon form factor $F_{1}$ at large~$Q^2$ 
}

\author{N.~Kivel\footnote{On leave of absence from Petersburg Nuclear Physics Institute, Gatchina, 188350, Russia}}
\affiliation{Institut f\"ur Kernphysik, Johannes Gutenberg-Universit\"at, D-55099 Mainz, Germany} 
\affiliation{Helmholtz Institut Mainz, Johannes Gutenberg-Universit\"at, D-55099 Mainz, Germany} 
\vskip1cm

\begin{abstract}

In  \cite{Kivel:2010ns}  we suggested  the  factorization formula for the nucleon form factors which consist of the sum of two 
contributions describing the  hard and soft  spectator scattering,  and  we provided a description of the soft rescattering contribution 
for the FF $F_{1}$ in terms of  convolution integrals of the  hard and hard-collinear coefficient functions with the appropriate
 soft matrix elements. 
 
In present  paper we investigate  the soft spectator scattering contribution for the FF $F_{1}$. 
We focus our attention on  factorization of the hard-collinear scale  $\sim Q\Lambda$  corresponding   
to transition from SCET-I to SCET-II.  
We compute the leading order  jet functions and find  that  the  convolution integrals over the soft fractions 
are logarithmically divergent. This  divergency is  the consequence of the boost invariance and does not 
 depend on the model of the soft correlation function describing the soft spectator quarks.  
 Using as example a two-loop diagram  we demonstrated  that such a divergency corresponds to the overlap of 
 the soft and collinear regions.  As a result one obtains  large rapidity logarithm  which  must be included 
 in the correct factorization formalism.

 We conclude   that a consistent description of the factorization for $F_{1}$  implies  the end-point collinear divergencies in the 
hard and soft spectator contributions, i.e. convolution integrals with respect to collinear fractions are not well-defined.   
Such scenario can only be realized   when the twist-3 nucleon  distribution amplitude  has  specific end-point behavior which  differs 
from  one expected from the evolution of the nucleon distribution amplitude.   Such behavior  leads to the violation of 
the collinear factorization for  the hard spectator scattering contribution.  We suggest that the soft spectator scattering 
and  chiral symmetry breaking  provide the  mechanism responsible for the violation of  collinear factorization in case of  
form factor $F_{1}$.
 
In spite of complexities of the SCET  factorization it  can be very  useful   for a phenomenological 
analysis of  hard exclusive reactions.  The basis for such approach is provided by  universality of the SCET-I form factors 
which can appear in different hard processes.    
We show that  using, so-called,  physical subtraction scheme SCET factorization in some cases allows to perform the systematical analysis 
of the hadronic processes  in the range of moderate values of $Q^{2}\sim 5-20$ GeV$^{2}$ where the hard collinear  
scale $\sim Q\Lambda$ is still not large

\end{abstract}


\maketitle

\section{Introduction}

 A substantial progress of the experimental  studies of the nucleon form factors (FFs) has been achieved during the last decade.  The polarization transfer method  allowed one  to measure accurately the proton FFs  up to momentum transfer $Q^2 \simeq 8.5$~GeV$^2$ \cite{Jones00, Punjabi:2005wq, Gayou02, Puckett:2010ac} , for recent reviews see, {\it e.g.},  
Refs.~\cite{HydeWright:2004gh,Arrington:2006zm,Perdrisat:2006hj}.   It also opened a possibility for  systematic studies of the FFs at large space-like $Q^{2}$  region in  the near future at the Jlab 12GeV  upgraded facility see, {\it e.g.},  \cite{Arrington:2011kb}.  At the same time the PANDA Collaboration at GSI is planning to carry
out precise measurements of the proton FFs at large time-like
momentum transfers, up to around 20 GeV$^2$,  
using the annihilation process $p+\bar{p}\rightarrow e^{+} + e^{-}$
\cite{Sudol:2009vc, Wiedner:2011mf}. These experiments will provide us with new information
on the FF behavior in the region of large momentum transfers that also provides  a strong motivation for the theoretical studies of the large-$Q$  behavior of the nucleon FFs.  

It is  known for a long time that in the case of nucleon FFs  the soft spectator scattering   is not  suppressed by inverse  powers of large $Q$   
\cite{Dun1980, Duncan:1979ny, Fadin1981, Fad1982}.  Moreover, results  of  different  phenomenological approaches \cite{Isgur:1984jm, Isgur:1988iw, Isgur:1989cy, Ioffe:1982qb, Nesterenko:1982gc, Braun:2001tj, Braun:2006hz}  
allow one to conclude  that in the region of  moderate  $Q^{2}\sim 5-15$GeV$^{2}$  such  mechanism plays the crucial role in order to obtain correct description of  the FFs data and also other hard processes with baryons see, {\it e.g.},  \cite{Radyushkin:1998rt, Diehl:2002yh, Kroll:2005ni, Feldmann:2011xf}.  
A systematic QCD approach for  description of  such soft mechanism is needed in order to develop  a model-independent formalism for description of  
existing and  future  experiments.

\begin{figure}[th]
\begin{center}
\includegraphics[height=1.5956in]{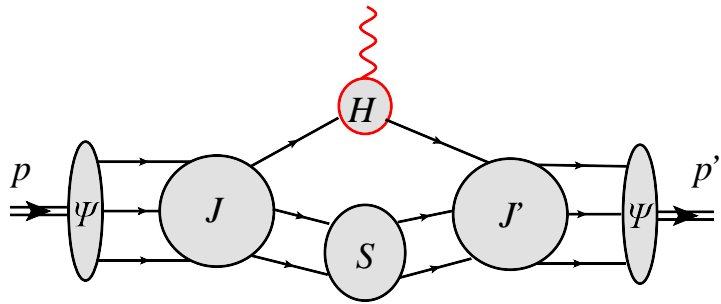}
\end{center}
\caption{Interpretation of the soft rescattering as a reduced diagram}%
\label{s-reduced-diagram}%
\end{figure}

In \cite{Kivel:2010ns}  the contribution of the soft spectator scattering   has been included into factorization scheme  using the  formalism of the soft collinear effective theory (SCET)\cite{Bauer:2000ew, Bauer2000, Bauer:2001ct, Bauer2001, BenCh, BenFeld03}. 
 The important feature in the description of the soft-overlap contribution is the  presence of the  hard-collinear scale $\sim Q\Lambda$. 
In order to take into account the soft rescattering mechanism one uses the two-step matching technique developed in SCET.  
Following from the leading-logarithmic analysis we suggested that the tentative factorization formula for  the Dirac FF $F_{1}$ can be written 
 as a sum of the hard and soft rescattering contributions:
 \begin{equation}
F_{1}\simeq F_{1}^{(h)}+F_{1}^{(s)},%
\label{F1:sum}
\end{equation}
where the hard rescattering part $F_{1}^{(h)}$ is  described by convolution of the hard coefficient function $\mathbf{H}$ with the nucleon distribution amplitudes (DAs) $\mathbf{\Psi}$:
\begin{equation}
F_{1}^{(h)}=\int Dx_{i}~\int Dy_{i}~\mathbf{\Psi}(y_{i})~\mathbf{H}(x_{i},y_{i}|Q)
~\mathbf{\Psi} (x_{i})\equiv\mathbf{\Psi}*\mathbf{H}*\mathbf{\Psi},
\label{F1:fact}%
\end{equation}

while the  soft contribution $F_{1}^{(s)}$  has the same scaling behavior  $\sim 1/Q^{4}$  as the hard spectator term $F_{1}^{(h)}$  
 and  can be presented in the following form~:
\begin{eqnarray}
F_{1}^{(s)} \simeq  
C_{A}(Q)
\int Dy_{i}\mathbf{\Psi}(y_{i})
 \int_{0}^{\infty}d\omega_{1}d\omega_{2}~\mathbf{J}'(y_{i},\omega_{i}Q)
 \int Dx_{i}\mathbf{\Psi}(x_{i}) 
\int_{0}^{\infty}d\nu_{1}d\nu_{2}~\mathbf{J}(x_{i},\nu_{i}Q)\boldsymbol{S}(\omega_{i},\nu_{i}).
\label{F1s:int}%
\end{eqnarray}
 This formula can be interpreted  graphically as reduced diagram in Fig.\ref{s-reduced-diagram}. 
The hard subprocesses in the soft spectator contribution are described by the hard coefficient function  $C_{A}$ and two 
hard-collinear jet functions $\mathbf{J}$, $\mathbf{J}^\prime$.  
They describe the parton scattering  with the hard  and hard-collinear momenta, respectively. 
The non-perturbative  DA $\mathbf{\Psi}$ and soft correlation function (SCF) $\boldsymbol{S}$ describe the long distance scattering of collinear and soft modes.   
The convolution integrals in Eq.~(\ref{F1s:int})
are performed with respect to the collinear fractions $x_{i}$ and $y_{i}$, and 
with respect to the soft spectator fractions $\omega_{i}, \nu_{i}\sim \Lambda$.  

In  \cite{Kivel:2010ns}  we  discussed  the matching of electro-magnetic current onto SCET-I operator, calculation hard coefficient function $C_{A}$ and 
resummation of Sudakov logarithms.  The  structure of the factorization formula (\ref{F1s:int}) was considered only qualitatively.  
In the current publication we present more  detailed consideration of this contribution. 
We shall  carefully consider the  matching of SCET-I to SCET-II and check  the validity  of the Eq.(\ref{F1s:int}).  

In \cite{Kivel:2010ns}  we assumed   the existence of all convolution integrals with respect to the collinear and soft fractions in   (\ref{F1s:int}). 
Such assumption is motivated by the following  observation.  First, one can easily observe that  collinear convolution integrals  in both contributions   
(\ref{F1:fact}) and   (\ref{F1s:int}) 
have a similar structure. Second,  if  one uses phenomenological models of  the nucleon DAs existing in literature  then the collinear integrals are well defined.  After that we expected that the integral over the soft fractions  must be also well defined.     
 However in this work we show  using  some general model-independent arguments  that  the convolution  with respect to the soft  fractions in  (\ref{F1s:int}) is divergent.  
 This divergence can be represented  as scaleless logarithmic integral 
 over one of the fractions: $\sim \int_{0}^{\infty} d\omega_{1}/\omega_{1}$.   Obviously, such integral has problems in both ultraviolet (UV) and infrared (IR) regions.  
 The divergence in UV-limit is a signal that  here one faces with the well known problem: overlapping of the soft and collinear modes.  
 
 In order to clarify the situation we perform more careful analysis of  the factorization (\ref{F1:sum}) in the perturbation theory using  perturbative interpretations for the DAs and soft correlation function. It turns out that  in the theory with massive quarks the collinear convolution integrals in  (\ref{F1:fact}) and   (\ref{F1s:int})  are also divergent.  
 This divergency arises due to the overlap  of the soft  and collinear sectors.  We demonstrate that there is  large rapidity logarithm $\sim \ln Q/\Lambda$  which can be  computed due to the overlap of  collinear and soft regions associated with    (\ref{F1:fact}) and   (\ref{F1s:int}). This is exactly the logarithm which was  computed in \cite{Dun1980}.  In order to compute   (\ref{F1:fact}) and  (\ref{F1s:int}) unambiguously  one has to define a certain prescription which allows one to separate the collinear and soft  sectors  and to avoid a double counting.  In the case of perturbation theory  such separation can be carried out using the the idea of subtractions  discussed  in QCD for the Sudakov FF \cite{Collins:1999dz} or, similar technique in SCET known as zero-bin subtraction method  \cite{Manohar:2006nz}.  
 We obtain that the perturbative  calculations of the studied diagram is in agreement with the factorization formula  (\ref{F1:fact}). 
However generalization of the perturbative results to realistic case faces with certain difficulties and  the description of the hard-collinear factorization even for  $F_{1}$ remains challenging.   

 The overlap of the collinear and soft regions imposes also a  qualitative idea  about  the end-point behavior of the twist-3 nucleon DA. 
  We  expect  that the end-point behavior of nucleon DA at  low normalization  differs from the behavior of the 
  asymptotic DA  $\mathbf{\Psi}_{as}(x_{i})\sim x_{1}x_{2}x_{3}$.  Usually the models of DA always vanish quadratically  when  the fractions $x_{i}$  are  small.   
 But  DA should have a different  behavior  in order to produce the end-point singularity in the convolution integrals. 
 We show  that  example of such behavior can be obtained even  in the perturbation theory with massive quarks:  
  for instance,  if $x_{1,2}\rightarrow 0$ and  $x_{1}\sim x_{2}$, then the perturbative DA  vanishes only linearly  as 
  $\mathbf{\Psi}(x_{1},x_{2}, x_{3}) \sim x_{1,2}$.  This already leads to the end-point  singularity in the collinear convolution integrals. 
   The presence of soft quark mass  in this calculation is necessary.  The combination of two possibilities: soft spectator scattering and 
   explicit chiral symmetry breaking due to the quark mass  leads to  the violation of the collinear factorization in case of helicity conserving Dirac form factor. 
  Taking into account the dynamical chiral symmetry breaking in QCD we expect a realization of the similar scenario for the nonperturbative nucleon DA.

 Our presentation is organized as follows. In Sec.~II we describe in details the soft rescattering contribution $F_{1}^{(s)}$  defined by 
 Eq.(\ref{F1s:int}), present the analytic results for the leading-order jet functions and discuss  the divergency of the soft convolution integrals.  Sec.~III is dedicated to the analysis of the overlap between the soft and collinear modes. Using two-loop QCD diagram we  calculate  the large rapidity  logarithm and  perform the interpretation of the  obtained results in terms of hard $F_{1}^{(h)}$ and soft  $F_{1}^{(s)}$  contributions. Generalizing these observations we suggest that collinear integrals in the factorization formulas for $F_{1}^{(h)}$ and  $F_{1}^{(s)}$  are also divergent and therefore the nucleon DA must have a specific end-point behavior. 
 In  Sec.IV we discuss the application of the SCET factorization for the phenomenological analysis at moderate values of $Q^{2}$.  In Sec.V, we summarize our results.

\section{ Soft  spectator scattering  contribution}
We  begin our discussion from the  description of the soft rescattering contribution. All notations are the same as in \cite{Kivel:2010ns} and are briefly described in Appendix A.   
For simplicity we consider proton  as the target nucleon.  In order to describe
$F_{1}^{(s)}$ we need to define  two  nonperturbative functions:  proton DA  $\Psi(x_{i})$ and  soft correlation function (SCF) $\boldsymbol{S}(\omega_{i},\nu_{i})$.

  The proton DA is a well-known object  defined as: 
\begin{equation}
~4\left\langle 0\left\vert W_{c}^{\dagger}[\lambda_{1}n]\xi^{u}_{\alpha_{1}}~W_{c}^{\dagger}\xi^{u}_{ \alpha_{2}}%
[\lambda_{2}n]~W_{c}^{\dagger}\xi^{d}_{\alpha_{3}}[\lambda_{3}n]\right\vert p\right\rangle
=\frac{\varepsilon^{ijk}}{3!}\int Dx_{i}~e^{-ip_{+}\left(  \sum x_{i}%
\lambda_{i}\right)  }\mathbf{\Psi}_{\bs \alpha}(x_{i}),
\label{DA:def}%
\end{equation}
where the measure reads$\ ~Dx_{i}=dx_{1}dx_{2}dx_{3}\delta(1-x_{1}-x_{2}-x_{3})$, $\varepsilon^{ijk}$ is the color factor, the multiindex  ${\bs \alpha}\equiv \alpha_{1}\alpha_{2}\alpha_{3} $,   and the index ``$c$'' denotes  in SCET-II the collinear fields :  
\begin{equation}
W^{\dagger}_{c}[x]\xi^{u}_{\alpha} \equiv\bar{ \text P}\exp\left\{  -ig\int_{-\infty}%
^{0}dt~(n\cdot A_{c})(x+tn)\right\} \left[\nbn  u_{c}(x)\right]_{\alpha}   ,
\end{equation}
 The function $\mathbf{\Psi}_{\bs \alpha}(x_{i})$ can be further represented as
\begin{eqnarray}
~\ \mathbf{\Psi}_{\bs \alpha}(x_{i}) & =V(x_{i})~p_{+}\left[  {\scriptstyle\frac{1}{2}}\Dslash{\bar n}~C\right]
_{\alpha_{1}\alpha_{2}}\left[  \gamma_{5}\nbn N\right]  _{\alpha_{3}}+A(x_{i})~p_{+}\left[
{\scriptstyle\frac{1}{2}}\Dslash{\bar n}\gamma_{5}C\right]
_{\alpha_{1}\alpha_{2}}\left[\nbn  N\right]  _{\alpha_{3}}\nonumber
\\  &  ~\ \ \ \ \ \ \ \ \ \ \ \ \ \ \ \ \ \ \ \ \ \ \ \ \ \ \ \ 
+T(x_{i})~p_{+}\left[  {\scriptstyle\frac{1}{2}}\Dslash{\bar n}\gamma_{\bot}~C\right]  _{\alpha_{1}\alpha_{2}}
\left[  \gamma^{\bot}\gamma_{5}\nbn N\right]  _{\alpha_{3}},
\label{Psi:def}
\end{eqnarray}
where $C$ is the charge conjugate matrix $C:~C\gamma_{\mu}C=\gamma_{\mu}^{T}$. The scalar functions $V,\, A,\, T$ depend on the collinear fractions $x_{i}$ and  from the factorization scale  which is not shown for simplicity.  Alternatively, these functions  can be represented through the one twist-3 DA  $\varphi_{N}(x_{i})$ \cite{Chernyak:1983ej}:
\begin{eqnarray}
&& 
V( x_{1}, x_{2}, x_{3})\equiv V(1,2,3)=\frac12(\, \varphi_{N}(1,2,3)+\varphi_{N}(2,1,3)\, ), 
\\ && 
A(1,2,3)=-\frac12 (\, \varphi_{N}(1,2,3)-\varphi_{N}(2,1,3)\, ), 
\\&&
 T(1,2,3)= \frac12(\, \varphi_{N}(1,3,2)+\varphi_{N}(2,3,1)\, ).
 \label{Tphi}
\end{eqnarray}
In many applications $\varphi_{N}(x_{i})$ usually   is  approximated by a few polynomials in $x_{i}$  with unknown coefficients. 
The polynomials represent the eigenfunctions  of the evolution kernels  and naturally arise from the solution of the evolution equation 
( see,{\it e.g.},  \cite{Braun:1999te, Braun:2003rp} and references there in).  
For example, the one of most popular parametrizations  for $\varphi_{N}(x_{i})$ can be represented as follows \cite{Chernyak:1983ej}:
\begin{eqnarray}
\varphi_{N}(x_{i})\simeq 120 x_{1}x_{2}x_{3} f_{N}(1+c_{10} (x_{1} - 2x_{2} + x_{3}) + c_{11}(x_{1} - x_{3})),
\label{phi}
\end{eqnarray}
The values of the nonperturbative constants $f_{N}$, $c_{i}$ are 
not important for our further discussion and we will not provide their numerical estimates.  

 A parametrization such as in Eq.(\ref{phi}) always fulfills  one  important property: it vanishes  quadratically at the boundary
 where the fractions are close to zero, for instance if $x_{1}\sim x_{2}\rightarrow 0$ one obtains
\begin{eqnarray}
\varphi_{N}(x_{1}, x_{2}, x_{3}) \sim x_{1} x_{2}.
\label{end-p}
\end{eqnarray}

Such behavior is an important requirement which ensures  the existence of the convolution integrals in the factorization formula (\ref{F1:fact}) because 
the hard coefficient functions, as a rule,  have the end-point singularities $\sim 1/x_{1}x_{2}$.  The end-point behavior (\ref{end-p}) can be associated  with the corresponding behavior 
of  the evolution kernel.  Performing the expansion of the DA $\varphi_{N}$ in terms of the eigenfunctions of the evolution kernel and neglecting  by the higher order harmonics  one 
always obtains the model which vanishes at the boundary  (\ref{end-p}).   However such approach is consistent  only if one assumes an appropriate convergence of the conformal expansion at a given normalization point.

The second non-perturbative input is the SCF  $\boldsymbol{S}$  introduced in Eq.(\ref{F1s:int}).  In SCET it is constructed from the soft quark and gluon fields $q$ and $A_{s}$, respectively.  The gluon fields enter only in the form of Wilson lines  such as
\begin{equation}
S_{\bar n}(x)=\text{P}\exp\left\{  ig\int_{-\infty}^{0}ds~\bar n\cdot A_{s}%
(x+s\bar n)\right\} ,\, \,
Y^{\dagger}_{n}(x)=\text{P}\exp\left\{  ig\int_{0}^{\infty}ds~n\cdot A_{s}%
(x+sn)\right\} 
\label{Ws}%
\end{equation}  
The definition of  $\boldsymbol{S}$ given in \cite{Kivel:2010ns} implies that it  is a tensor with respect to Dirac indices. 
It is convenient to  rewrite it in terms of scalar functions. This can be done with the help of Fierz transformation so that the result reads:
\begin{eqnarray}
\boldsymbol{S}(\omega_{i},\nu_{i})=\frac{1}{8}C\Dslash{n}\otimes \frac{1}{8}\Dslash{\bar n}C\,  S_{V}(\omega_{i},\nu_{i})
+\frac{1}{8}C\Dslash{n}\gamma_{5}\otimes \frac{1}{8}\Dslash{\bar n}\gamma_{5}C\,  S_{A}(\omega_{i},\nu_{i})+
\frac{1}{4}C\Dslash{n}\gamma_{\bot}^{\mu}\otimes \frac{1}{4}\Dslash{\bar n}\gamma_{\bot}^{\nu}C\, S^{\mu\nu}_{T}(\omega_{i},\nu_{i}).
\end{eqnarray}
The scalar functions $S_{V,A,T}$ defined as 
\begin{equation}
~S_{X}(\omega_{i},\nu_{i})=\int\frac{d\eta_{1}}{2\pi}%
\int\frac{d\eta_{2}}{2\pi}~e^{-i\eta_{1}\nu_{1}-i\eta_{2}\nu_{2}}\int
\frac{d\lambda_{1}}{2\pi}\int\frac{d\lambda_{2}}{2\pi}e^{i\lambda_{1}%
\omega_{1}+i\lambda_{2}\omega_{2}}\left\langle 0\right\vert O_{X}(\eta_{i},\lambda_{i})\left\vert 0\right\rangle ,
\label{Sdef}%
\end{equation}
with the operators
\begin{eqnarray}
O_{X}(\eta_{i},\lambda_{i}) & =&  \varepsilon^{i^{\prime}j^{\prime
}k^{\prime}}
\left[  Y_{n}^{\dag}(0)\right]  ^{i^{\prime}l}
\left[  Y_{n}^{\dag}q(\lambda_{1}n)\right]^{j^{\prime}} C\Gamma_{X}\left[Y_{n}^{\dag}q(\lambda_{2}n)\right] ^{k^{\prime}} 
\nonumber \\&  \times & 
\varepsilon^{ijk}\left[  S_{\bar{n}}(0)\right]  ^{li}
\left[  \bar{q} S_{\bar{n}}(\eta_{1}\bar{n})\right]^{j} \bar\Gamma_{X}C
\left[\bar{q}S_{\bar{n}}(\eta_{2}\bar{n})\right]^{k},
\label{opOX}
\end{eqnarray}
where 
\begin{eqnarray}
\{ \Gamma_{X}\otimes \bar\Gamma_{X}\}_{X=V,A,T}=\{\Dslash{n}\otimes\Dslash{\bar n},\Dslash{n}\gamma_{5}\otimes \Dslash{\bar n}\gamma_{5},
\Dslash{n}\gamma_{\bot}\otimes\Dslash{\bar n}\gamma_{\bot}  \}.
\label{VAT}
\end{eqnarray}

The flavor structure of the SCFs $S_{X}$ is defined by the flavor structure of the proton: it can be described  as either $uu$- or $ud$-combinations. Below we will show  that 
the tensor component $S_{T}$ does not contribute to $F_{1}$ and hence we can conclude that the matrix   $\boldsymbol{S}(\omega_{i},\nu_{i})$ is presented by 
four scalar functions: $S_{A,V}^{\text{uu,ud}}(\omega_{i},\nu_{i})$.   Indeed, SCFs also depend on the factorization scale, which is  not shown for simplicity. 

One has to keep in  mind that the arguments   $\omega_{i}$ and $\nu_{i}$ (soft fractions) are of order $\Lambda$.  Obviously, these variables  can be associated with the light-cone 
projections of the soft spectators momenta which are defined to be positive.  For instance,  the leading order calculation of the SCF $S_{V}^{\text{ud}}(\omega_{i},\nu_{i})$ 
in perturbation theory   yields (see details in Appendix B)
\begin{eqnarray}
\left[S_{V}^{\text{ud}}(\omega_{i},\nu_{i}) \right] _{\text{LO}}=\frac{3m^{2}}{16\pi ^{6}}~\theta (\nu _{i}>0)
\theta (\omega _{i}>0)~~\theta (\omega _{1}\nu _{1}>m^{2})\theta (\omega _{2}\nu _{2}>m^{2}),
\label{SLO}
\end{eqnarray}
where $m$ denotes the mass of  soft quarks.  Notice that $S_{V,A}$ is proportional to the  square of mass $m^{2}$ which arises from the numerators of the soft quark propagators and this fact, as we will see later, has an important consequence.

The last elements in Eq.(\ref{F1s:int}) which one has to introduce are  jet functions $\mathbf{J}$ and $\mathbf{J}'$.  These are  hard-collinear coefficient functions which 
can be computed in pQCD if the hard-collinear scale $Q\Lambda$ is quite large. They  appear in the matching of hard-collinear modes onto collinear and soft fields in SCET-II. Jet functions can be computed 
from the $T$-products which schematically can be written as \cite{Kivel:2010ns}:
\begin{align}
 T\left(  \bar{\xi}_{hc}^{\prime}W^{\prime}~e^{i\mathcal{L}^{({n})}_{\text{SCET-I }}%
}~\right)\simeq 
T\left(  \bar{\xi}_{hc}^{\prime}W^{\prime},{i\mathcal{L}^{({1})}_{\xi' q}}, {i\mathcal{L}^{({1})}_{\xi' q}}, {i\mathcal{L}^{({0})}_{\xi' \xi'}}
~\right)
 =\mathbf{~~}\bar{\xi}_{c}^{\prime}\bar{\xi}_{c}^{\prime}%
\bar{\xi}_{c}^{\prime}\ast\mathbf{J}_{LO}^{\prime}\ast qq~, \label{Jdefout}
\\  
T\left( W^{\dag} {\xi}_{hc}~e^{i\mathcal{L}^{(\bar{n})}_{\text{SCET-I}}%
}~\right)  
\simeq 
T\left(  W^{\dag} {\xi}_{hc}, {i\mathcal{L}^{({1})}_{\xi q}}, {i\mathcal{L}^{({1})}_{\xi q}}, {i\mathcal{L}^{({0})}_{\xi \xi}}
~\right)
= \bar q\bar q\ast \mathbf{J}_{LO}\ast{\xi}_{c}{\xi}_{c}{\xi}_{c}~,\ \ \ \label{Jdef}%
\end{align}
where the asterisks denote the convolutions with respect to collinear and soft fractions.  The $T$-products which  must be computed in SCET-I are shown explicitly, 
$\mathcal{L}^{({0})}_{\xi \xi}$ and $\mathcal{L}^{({1})}_{\xi q}$ denote the leading and subleading terms in the SCET-I Lagrangian, respectively.  
To the leading order accuracy  we did not find any other  combinations of the subleading terms which can describe the contributions with the soft spectators, in particular,
the contribution with  one soft spectator. 

\begin{figure}[th]
\begin{center}
\includegraphics[
natheight=0.531900in,
natwidth=6.103000in,
height=0.5319in,
width=6.103in
]{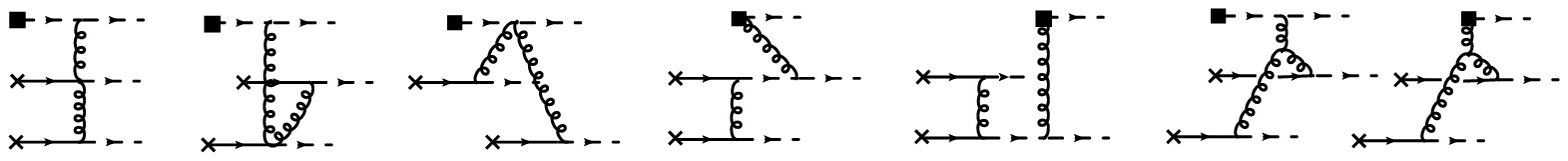}
\end{center}
\caption{ Leading order SCET diagrams required for the calculation of jet
functions. The inner dashed and curly lines denote hard-collinear quarks and
gluons, external dashed lines correspond to collinear quarks, fermion lines 
with crosses denote soft quarks. Black square denotes the vertex of
the SCET-I operator. }%
\label{jet-f-diagrams}%
\end{figure}

The matrix elements  of the collinear operators in Eq.(\ref{Jdef}) yield twist-3 DAs (\ref{DA:def}) describing  the initial and final protons, 
the soft fields $q$ are combined into SCF $\boldsymbol{S}$.  
Because the collinear operators in the both equations in (\ref{Jdef}) are the same, the functions  $\mathbf J$ and $\mathbf J'$  are also the same, therefore one needs  to compute only one of it. 
Corresponding leading order diagrams are shown in Fig.\ref{jet-f-diagrams}.
The last two diagrams with the three gluon vertex have
vanishing color factors and therefore do not contribute. This is in  agreement
with the  observation made in Ref.~\cite{Fadin1981}.  The soft-collinear vertices  in the diagrams is obtained from the subleading SCET Lagrangian ${\mathcal L}^{(1)}_{\xi q}$, see details in \cite{Kivel:2010ns}.  
Computing these diagrams we projected the Dirac indices according to our  definitions of the DA  (\ref{Psi:def}) and SCF (\ref{opOX}).  
We find that the  contribution with $S_{T}$  corresponding to the tensor projection in eq. (\ref{VAT}) has trivial coefficient function.  This result is not an accident and can be explained by the angular momentum conservation  (quite similar to  discussion in  \cite{Collins:1999un}). Therefore we expect that corresponding hard kernel is trivial  to all orders in $\alpha_{s}$.

 In order to present  the results for the jet functions let us rewrite Eq.(\ref{F1s:int})   in the following form:
\begin{equation}
~F_{1}^{(s)}(Q)=C_{A}(Q)~\{\, e_{u}~f_{1}^{\text{u}}(Q)+e_{d}~f_{1}^{\text{d}}(Q) \},
\label{f1ud}
\end{equation}
where $e_{u,d}$ denote quark charges.  The hard coefficient function $C_{A}(Q)$ in Eq.(\ref{f1ud}) includes all large logarithms so that  SCET-I form factors $f_{1}^{\text{u,d}}$  depend only from the hard-collinear scale $Q\Lambda$ and  defined as 
\begin{equation}
\left\langle p^{\prime}\right\vert 
 \bar {\xi}_{hc}^{\prime\, q }W^{\prime} \gamma_{\bot}^{\mu}~W^{\dag} \xi^{q}_{hc}
\left\vert p \right\rangle_{\text{SCET} }
=\bar{N}(p')\frac{ \Dslash{\bar{n}}\Dslash{ n} }{4}\gamma_{\bot}^{\mu}N(p)~f_{1}^{q}(Q, \mu_{hc} \simeq Q\Lambda)\equiv  f_{1}^{q}(Q),
\label{f1q:def}
\end{equation}
where the index $q$ describe the flavor of the hard-collinear field (active quark).  Computing the diagrams in Fig.\ref{jet-f-diagrams}  we obtained analytical  expressions 
for the leading order jet functions. Our results  can be presented in the following way 
\begin{equation}
~f_{1}^{\text{u}}(Q)=
 I_{A}^{\text{u}}(\omega_{1},\omega_{2})\ast
  S_{A}^{\text{ud}}(\omega_1,\omega_2,\nu_{1},\nu_{2})\ast
  I_{A}^{\text{u}}(\nu_{1},\nu_{2})
  -I_{V}^{\text{u}}(\omega_{1},\omega_{2})\ast
   S_{V}^{\text{ud}}(\omega_1,\omega_2,\nu_{1},\nu_{2})\ast
   I_{V}^{\text{u}}(\nu_{1},\nu_{2})  ,
   \label{f1u}
\end{equation}
\begin{equation}
~f_{1}^{\text{d}}(Q)=
I_{A}^{\text{d}}(\omega_{1},\omega_{2})\ast  {S}_{A}^{\text{uu}}(\omega_1,\omega_1,\nu_{1},\nu_{2})\ast  I_{A}^{\text{d}}(\nu_{1},\nu_{2})
- I_{V}^{\text{d}}(\omega_{1},\omega_{2})\ast {S}_{V}^{\text{uu}}(\omega_1,\omega_2,\nu_{1},\nu_{2})\ast I_{V}^{\text{d}}(\nu_{1},\nu_{2})   ,
\label{f1d}
\end{equation}
where asterisk  denotes the convolution integral with respect to the soft fractions, for instance
\begin{equation}
 I_{A}^{\text{d}}(\omega_{1},\omega_{2})\ast  {S}_{A}^{\text{uu}}(\omega_1,\omega_1,\nu_{1},\nu_{2})
 =\int_{0}^{\infty} d\omega_{1} d\omega_{2} \,
 I_{A}^{\text{d}}(\omega_{1},\omega_{2})  {S}_{A}^{\text{uu}}(\omega_1,\omega_1,\nu_{1},\nu_{2}).
\end{equation}
The proton DAs and integrations over the collinear fractions enter into the  functions $I_{V,A}^{\text{u,d}}$ in Eqs. (\ref{f1u},\ref{f1d}).  These functions  read:
\begin{eqnarray}
I_{V}^{\text{u}}(\omega_1,\omega_2) & =&   (V-A-2T)(y_{1},y_{2},y_{3})\ast J_{a}^{\text{u}}  (y_{1},y_{2},y_{3},\omega_{1},\omega_{2}) 
 \nonumber \\
&+&(A-V-2T)(y_{1},y_{2},y_{3})\ast J_{b}^{\text{u}} (y_{1},y_{2},y_{3},\omega_{1},\omega_{2}),
\label{JVu}
\end{eqnarray}
\begin{eqnarray}
I_{A}^{\text{u}}(\omega_1,\omega_2) & =&   (V-A-2T)(y_{1},y_{2},y_{3})\ast J_{b}^{\text{u}}  (y_{1},y_{2},y_{3},\omega_{1},\omega_{2}) 
 \nonumber \\
&+&(A-V-2T)(y_{1},y_{2},y_{3})\ast J_{a}^{\text{u}} (y_{1},y_{2},y_{3},\omega_{1},\omega_{2}),
\label{JAu}
\end{eqnarray}
\begin{eqnarray}
I_{V}^{\text{d}}(\omega_1,\omega_2)  =   V(y_{1},y_{2},y_{3})\ast J_{a}^{\text{d}}  (y_{1},y_{2},y_{3},\omega_{1},\omega_{2}) 
-A(y_{1},y_{2},y_{3})\ast J_{b}^{\text{d}} (y_{1},y_{2},y_{3},\omega_{1},\omega_{2}),
\label{JVd}
\end{eqnarray}
\begin{eqnarray}
I_{A}^{\text{d}}(\omega_1,\omega_2)  =   
 -A(y_{1},y_{2},y_{3})\ast J_{a}^{\text{d}} (y_{1},y_{2},y_{3},\omega_{1},\omega_{2}),
\label{JAd}
\end{eqnarray}
where  the asterisk again denotes the convolution integral over collinear fractions: 
\begin{equation}
V(y_{1},y_{2},y_{3})\ast J^{\text{d}}  (y_{1},y_{2},y_{3},\dots)=\int Dy_{i}V(y_{1},y_{2},y_{3})J^{\text{d}}  (y_{1},y_{2},y_{3},\dots).
\end{equation}
 The leading order hard-collinear jet-functions $J_{a,b}^{\text{u,d}}$  in Eqs.(\ref{JVu}-\ref{JAd}) read ($\bar y_{i}=1-y_{i}$):
\begin{eqnarray}
J_{a}^{\text{u}} (y_{1},y_{2},y_{3};~\omega_{1},\omega_{2})&  =&\alpha^{2}_{s}(\mu_{hc})  \frac{ 8\pi^{2} } {27}\,  
\frac{1}{\left(  \omega_{1}+\omega_{2}\right)  \omega_{1}\omega_{2}}
\frac{1}{y_{2}y_{3}}
\left( 
\frac{1}{\bar y_{2}}+\frac{1}{\bar y_{3}}
-
\frac{4( \omega_{1} y_{2}+\omega_{2}y_{3})  }
{ \bar{y}_{1}^{2}( \omega_{1}+\omega_{2} )}
\right),
\label{Jua}
\\
J_{b}^{\text{u}}(y_{1},y_{2},y_{3};~\omega_{1},\omega_{2}) &  =&\alpha^{2}_{s}(\mu_{hc})  \frac{ 8\pi^{2} } {27}\,  \frac{1}{\left(  \omega_{1}+\omega_{2}\right)  \omega
_{1}\omega_{2}} \frac{1}{\bar{y}_{2}\bar{y}_{3}}
\left(\frac{1}{y_{3}}-\frac{1}{y_{2}} \right),
\end{eqnarray}%
and
\begin{eqnarray}
J_{a}^{\text{d}}(y_{1},y_{2},y_{3};~\omega_{1},\omega_{2}) &  =&
\alpha^{2}_{s}(\mu_{hc})  \frac{ 8\pi^{2} } {27}\,  
\frac{1}{\left(  \omega_{1}+\omega_{2}\right)  \omega_{1}\omega_{2}}\frac{1}{y_{1}y_{2}}
\left(  
\frac{1}{\bar{y}_{1}}+\frac{1}{\bar{y}_{2}}
-
\frac{4(\omega_{1}y_{1}+\omega_{2}y_{2})}{\bar y_{3}^{2}(\omega_{1}+\omega_{2})}
\right)  ,
\\
J_{b}^{\text{d}}(y_{1},y_{2},y_{3};~\omega_{1},\omega_{2}) &  =&
\alpha^{2}_{s}(\mu_{hc})  \frac{ 8\pi^{2} } {27}\,  
\frac{1}{\left( \omega_{1}+\omega_{2}\right)  \omega_{1}\omega_{2}}~\frac{1}{\bar{y}_{1}
\bar{y}_{2}}\left(  \frac{1}{y_{2}}-\frac{1}{y_{1}}\right)  .
\label{Jdb}
\end{eqnarray}
The  argument of the QCD running coupling is defined by the hard-collinear scale:  $\mu_{hc}\sim Q\Lambda$. 
From results  (\ref{f1u})-(\ref{f1d})  one can deduce that the convolution integrals in (\ref{JVu})-(\ref{JAd}) with  the proton DAs vanishing at the boundary as in Eq.(\ref{end-p}) are well defined.  
Therefore one can assume that  in the absence of other dominant regions, the convolution integrals 
with respect to  soft fractions in (\ref{f1u})-(\ref{f1d}) should  also  be finite.  One can expect that the  SCFs are concentrated in the region where the soft fractions is of order $\Lambda$ and fall quickly  in the region where the soft fractions are much larger than $\Lambda$.

However the assumption of convergence is not  correct.  Using  the perturbative expression (\ref{SLO}) instead of  nonperturbative function one can easily obtain, that any convolution integral with respect to the soft fractions is  logarithmically divergent and proportional to $\sim \ln [\mu_{UV}/\mu_{IR}]$ where scale $\mu_{UV}$ and $\mu_{IR}$ represent the UV and IR cut-off respectively.  This is exactly the  logarithmic contribution  which was found in \cite{Duncan:1979ny}  and later studied  in \cite{Fadin1981}  and called as 
``nonrenormalization group type logarithmic contribution''.

 Using the leading order  expressions  given in Eqs.(\ref{Jua})-(\ref{Jdb}) one can show that the similar situation also takes place  for the nonperturbative SCFs.
 Recall that the soft fractions $\omega_{i}$ and $\nu_{i}$ can  be associated with  plus and minus projections of the soft momenta describing soft spectators. 
Therefore  the boost invariance implies  that SCF depends on the products $\omega_{i}\nu_{j}$:
\begin{equation}
S_{X}(\omega_{1},\omega_{2},\nu_{1},\nu_{2})
\equiv 
S_{X}(\omega_{1}\nu_{1},\omega_{1}\nu_{2},\omega_{2}\nu_{2},\omega
_{2}\nu_{1}), 
\label{S-boost}%
\end{equation}  
Using this observation one can easily obtain  that the convolution integrals
of jet-functions and SCF in Eqs.(\ref{f1u})-(\ref{f1d}) are divergent. Consider, as example, the integral  from the  Eq.(\ref{f1u}):
\begin{eqnarray}
J&=&
I_{A}^{\text{u}}(\omega_{1},\omega_{2})\ast
  S_{A}^{\text{ud}}(\omega_1\nu_{1},\omega_1 \nu_{2} , \omega_2\nu_{1},\omega_2\nu_{2})\ast
  I_{A}^{\text{u}}(\nu_{1},\nu_{2}),
\end{eqnarray} 
Using substitutions
\begin{eqnarray}
\nu_{1}=\nu_{1}^{\prime}/\omega_{1},~\nu_{2}=\nu_{2}^{\prime}/\omega_{1},\ \omega_{2}=\omega_{2}^{\prime}\omega_{1}
\end{eqnarray}
one obtains
\begin{eqnarray}
J=\int_{0}^{\infty}\frac{d\omega_{1}}{\omega_{1}}\int_{0}^{\infty}d\omega_{2}^{\prime}
\int_{0}^{\infty}d\nu_{1}^{\prime}\int_{0}^{\infty}d\nu_{2}^{\prime}\, 
I_{A}^{\text{u}}(\omega_{1},\omega_{2}^{\prime}\omega_{1})
I_{A}^{\text{u}}(\nu_{1}^{\prime}/\omega_{1},\nu_{2}^{\prime}/\omega_{1})
  S_{A}^{\text{ud}}(\nu_{1}^{\prime},\nu_{2}^{\prime} , \omega_2^{\prime}\nu_{1}^{\prime},\omega_2^{\prime}\nu_{2}^{\prime}).  
\end{eqnarray} 
Using homogeneity of the leading order jet functions in  Eqs.(\ref{Jua})-(\ref{Jdb}) yields
 \begin{eqnarray}
 I_{A}^{\text{u}}(\omega_{1},\omega_{2}^{\prime}\omega_{1})
I_{A}^{\text{u}}(\nu_{1}^{\prime}/\omega_{1},\nu_{2}^{\prime}/\omega_{1})=
I_{A}^{\text{u}}(1,\omega_{2}^{\prime})
I_{A}^{\text{u}}(\nu_{1}^{\prime},\nu_{2}^{\prime})
 \end{eqnarray}
and one obtains
\begin{eqnarray}
J=\int_{0}^{\infty}\frac{d\omega_{1}}{\omega_{1}}\int_{0}^{\infty}d\omega_{2}^{\prime}
\int_{0}^{\infty}d\nu_{1}^{\prime}\int_{0}^{\infty}d\nu_{2}^{\prime}\, 
I_{A}^{\text{u}}(1,\omega_{2}^{\prime})
I_{A}^{\text{u}}(\nu_{1}^{\prime},\nu_{2}^{\prime})
  S_{A}^{\text{ud}}(\nu_{1}^{\prime},\nu_{2}^{\prime} , \omega_2^{\prime}\nu_{1}^{\prime},\omega_2^{\prime}\nu_{2}^{\prime}).
  \label{JS}
  \end{eqnarray} 
 The first integral in the last Eq.(\ref{JS}) is divergent $\int_{0}^{\infty}\frac{d\omega_{1}}{\omega_{1}}\sim \ln [\mu_{UV}/\mu_{IR}]$.
 The similar arguments can be used for the all contributions in Eqs.(\ref{f1u})-(\ref{f1d}).  
 Then, for instance,  for the $u$-quark form  factor in Eq.(\ref{f1u}) one obtains
 \bea
 f_{1}^{\text{u}}(Q)= \int_{0}^{\infty}\frac{d\omega_{1}}{\omega_{1}}
 \int_{0}^{\infty}d\omega_{2}^{\prime}
\int_{0}^{\infty}d\nu_{1}^{\prime}\int_{0}^{\infty}d\nu_{2}^{\prime}\, 
\left\{
I_{A}^{\text{u}}(1,\omega_{2}^{\prime})
I_{A}^{\text{u}}(\nu_{1}^{\prime},\nu_{2}^{\prime})
  S_{A}^{\text{ud}}(\nu_{1}^{\prime},\nu_{2}^{\prime} , \omega_2^{\prime}\nu_{1}^{\prime},\omega_2^{\prime}\nu_{2}^{\prime})
\right. \\ \nonumber  
\left. -I_{V}^{\text{u}}(1,\omega'_{2}) I_{V}^{\text{u}}(\nu'_{1},\nu'_{2})
   S_{V}^{\text{ud}}(\nu'_1,\nu'_2,\omega_2^{\prime}\nu'_{1},\omega_2^{\prime}\nu'_{2})
    \right\},
   \label{f1udiv}
 \eea
 The the coefficient in front of divergent factor depends on the different combinations of different nonperturbative  functions and 
 we do not see any arguments which allows one to conclude that this coefficient may vanish.  
 Hence we can conclude that the soft convolution integrals in the soft spectator contribution are divergent and this singularity is  
 independent on the particular properties of SCF.   The same observation can also be done  for the $d$-quark form  factor in Eq(\ref{f1d}). 
 
 In conclusion, we have shown by the explicit leading order calculation of the jet functions  that the soft spectator contribution is not well defined. 
 The convolution integrals with respect to the soft fractions are divergent and these, so-called,  end-point singularities do not cancel in the sum of all diagrams.    
 
Is this divergency a signal that the factorization scheme suggested in Eqs.(\ref{F1:sum}-\ref{F1s:int}) is incomplete? 
The form of the divergent contribution allows one  to assume that dimensionless integral $\int_{0}^{\infty}\frac{d\omega_{1}}{\omega_{1}}$ 
indicates that the soft region overlaps with the collinear one as it happens, for instance, in the classical case of Sudakov form factor \cite{Collins:1999dz} . 
Then we may also expect that  contributions associated with the collinear  regions overlap with the soft region. If such situation takes place then we  
must observe IR-divergencies in the collinear convolution integrals which describe the hard rescattering term and  also referred   as   end-point divergencies.  

However,  the corresponding collinear integrals computed with the existing models of DA, like one in Eq.(\ref{phi}),  are well defined.  At first glance such
situation may look as contradictory, however it is not like that. In order to explain this observation one has to take into account following arguments.  

First, the end-point behavior of the DA models  Eq.(\ref{phi}) is closely associated with the end-point behavior of the QCD evolution.
In particular, the asymptotic DA is the eigenfunction of the evolution kernel with the smallest eigenvalue which is the anomalous 
 dimension of the corresponding three quark operator. 

The second  observation is given by consideration of the soft correlation function in perturbation theory.  
We obtained  that the first nontrivial contribution  is associated with the masses of the soft spectators, see Eq.(\ref{SLO}).
On the other hand for the calculation of the evolution kernel the  mass of the quarks is not needed and the chiral symmetry in perturbation theory is not broken. 
The terms proportional to the masses in such calculation are finite and as a rule neglected.  But for the calculation of the soft spectator contribution in 
perturbation theory the presence of  the soft spectator quark masses is necessary, see details in Appendix B.  
The introduction of mass for the soft spectator quarks in the perturbative calculations is directly related with the chiral symmetry breaking in pQCD.  
Then we may conclude that the mechanism of the violation of the factorization  in our case is closely associated with chiral 
symmetry breaking and therefore it can not be observed if this symmetry is preserved. Such scenario may explain why we do not observe any 
end-point singularities in the collinear integrals  computed with the models like  one in Eq.(\ref{phi}) for which 
the end-point behavior is strongly motivated by  QCD evolution. 

In order to verify this picture and to clarify the mechanism of the collinear and soft overlap in QCD  we  suggest to perform the analysis of  the factorization
 in the perturbative sector  with massive quarks.  In this case the mass plays the role of the soft scale and 
  we can unambiguously define  the all soft matrix elements in the perturbation theory.  
In he collinear sectors  this allows us to compute not only the logarithmic contributions associated with the evolution of DA 
 but also the finite contributions  which can appear due to quark mass.     

In the next section we  present investigation of  the factorization properties in pQCD using as example the 2-loop  
diagram which has both contributions associated with  the soft and collinear  sectors.   
 
\section{ Soft-collinear overlap and separation of the hard and soft  spectator scattering contributions }
\label{secIII}

\subsection{ Overlap of the soft and collinear regions }
\label{secIIIA} 
\begin{figure}[th]
\begin{center}
\includegraphics[
height=.8in
]{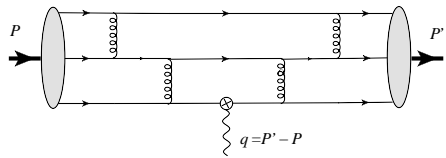}
\end{center}
\caption{ Two-loop diagram which obtain contributions from the hard, collinear and soft regions.}%
\label{2-loop-diagram}%
\end{figure}

In order to perform the required  analysis we  consider the diagram shown in Fig.\ref{2-loop-diagram}. This diagram  was discussed in 
\cite{Kivel:2010ns}   in order to show nontrivial contribution  arising from the soft region.  But we did not investigate the possibility of the overlap of the soft region  with  other dominant regions. We consider this more accurately now because it will help us to solve the problem with the soft divergent integral and understand the factorization for $F_{1}$. 
The contribution of this diagram into the nucleon FF $F_{1}$ can be written as 
\bea
\bar N' \gamma^{\mu}_{\bot}N F_{1}[D]= \int Dx_{i}\int Dy_{i}\,  \bs{\Psi}_{\bs \alpha}(y_{i})
\left[ D^{\mu_{\bot}} (x_{i}, y_{i}) \right]_{\bs {\alpha \beta} } 
 \bs{\Psi}_{\bs \beta}(x_{i})
\eea
where DA $ \bs{\Psi}$ is associated with the blobs in Fig.\ref{2-loop-diagram}.  In order to simplify our consideration we use simple observation which follow from Eq.(\ref{Psi:def}):
\be
\bs{\Psi}(y_{i})\Dslash{n}=0,\, \,   \Dslash{\bar n}\bs{\Psi}(x_{i})=0.
\label{LTP}
\ee 
This allows us in the intermediate calculations  to substitute instead of  full DA $\bs{\Psi}$ the large components of  collinear spinors:
\bea
 \bs{\Psi}_{\bs \beta}(x_{i})\rightarrow \left[\nbn u(p_{1})\right]_{\beta_{1}}\left[\nbn u(p_{2})\right]_{\beta_{2}}\left[\nbn d(p_{3})\right]_{\beta_{3}}
 \equiv [\xi_{1}]_{\beta_{1}}[\xi_{2}]_{\beta_{2}}[\xi_{3}]_{\beta_{3}}\equiv \chi_{\bs \beta},
 \label{collPin}
 \\
 {\bs \Psi}_{\bs \alpha }(y_{i})\rightarrow \left[ \bar u(p'_{1})\nbn\right]_{\alpha_{1}} \left[\bar u(p'_{2})\nbn \right]_{\alpha_{2}}
 \left[\bar d(p'_{3})\nbn \right]_{\alpha_{3}} 
 \equiv
  [\bar \xi^{\prime}_{1}]_{\alpha_{1}}[\bar \xi^{\prime}_{2}]_{\alpha_{2}}[\bar \xi^{\prime}_{3}]_{\alpha_{3}}\equiv \bar \chi'_{\bs \alpha},
  \label{collPout}
\eea

 For simplicity we always assume that the bottom fermion line in the diagram corresponds to $d$-quark. 
 We always assume that color indices are properly contracted and don't show them explicitly.  In order to simplify our consideration we will not contract the Dirac 
 indices and compute corresponding traces. 
 It is also clear that relation of the perturbative expressions with the corresponding realistic amplitudes  with nonperturbative DAs can be obtained 
 by  re-substitution  $\chi\rightarrow \bs \Psi$. 
   
  Therefore the  contribution of the diagram in Fig.\ref{2-loop-diagram} can be associated 
 with the perturbative diagram in Fig.\ref{2-loop-D} and corresponding analytical expression reads:
\begin{figure}[th]
\begin{center}
\includegraphics[
height=1.5in
]{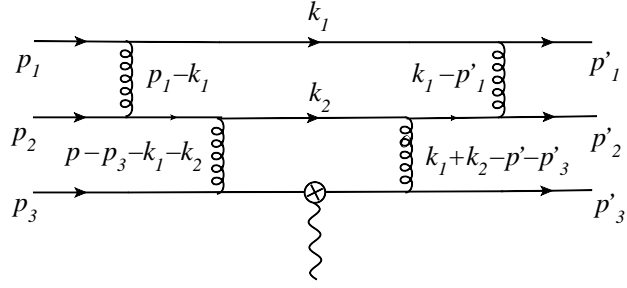}
\end{center}
\caption{Two-loop diagram with momenta flow. }%
\label{2-loop-D}%
\end{figure}
\bea
{D^{\mu}}&=&
~\mathcal{C} \int d^4 k_{1} d^4 k_{2}\frac{1}{\left[  k_{1}^{2}-m^{2}\right]  \left[  k_{2}^{2}-m^{2}\right]  } 
 \nonumber\\[0.5mm]
&\times& \frac{~\bar\xi_{1}^{\prime}\gamma^{i}\left( \hat k_{1}+m\right)  \gamma^{j}\xi_{1}~ ~
\bar\xi_{2}^{\prime}
\gamma^{i}(\hat p^{\prime}-\hat p_{3}^{\prime}-\hat k_{1}+m)\gamma^{\alpha}(\hat k_{2}+m)\gamma^{\beta}
(\hat p-\hat p_{3}-\hat k_{1}+m)\gamma^{j}~\xi_{2} }
{ [\left(  p-p_{3}-k_{1}\right)^{2}-m^{2}]     [\left(  p^{\prime}-p_{3}^{\prime}-k_{1}\right) ^{2}-m^{2}] 
\left(  k_{1}+k_{2}-p^{\prime}+p_{3}^{\prime}\right) ^{2}    \left(  p - p_{3} - k_{1} - k_{2} \right)^{2} }
\nonumber\\[0.5mm]
 &\times & \frac{     
\bar \xi_{3}^{\prime}~\gamma^{\alpha}
\left( \hat p^{\prime}-\hat k_{1}-\hat k_{2}+m\right) \gamma^{\mu} \left(  \hat p-\hat k_{1}-\hat k_{2}+m \right)  
\gamma^{\beta}~\xi_{3}~} 
{  [\left(  p^{\prime}-k_{1}-k_{2}\right)^{2}-m^{2}]  [\left(  p-k_{1}-k_{2}\right) ^{2}-m^{2}] 
\left(  p_{1}-k_{1}\right)^{2} \left( k_{1}-p_{1}^{\prime}\right) ^{2}   }
\label{D}%
\eea
The quark mass $m$ is used as a soft scale in order to regularize  IR-divergencies and to describe the soft contribution, see Eq.(\ref{SLO}).  
The factor $\mathcal{C}$ accumulates color structures and others factors according to Feynman rules.     

Our task is to clarify the relation between the collinear and soft regions performing the interpretation of $D^{\mu}$ in terms of different contributions according to factorization formulas (\ref{F1:fact})  and  (\ref{F1s:int}). If the overlap of these sectors  is not possible then we may be find  the missing elements in the factorization description. Notice that  diagram in 
Fig.\ref{2-loop-D} does not have UV-divergent subgraphs and therefore we do not need to consider renormalization of QCD parameters like quark mass and running coupling.  Taking into account the logarithmic structure of the hard spectator term  $F_{1}^{(h)}$  we may expect that after 2-loop integration  the answer   can be schematically written  as 
\be
D^{\mu_{\bot}}=\frac{1}{Q^{6}} [A^{\mu_{\bot}} \ln^{2} Q^{2}/m^{2}+B^{\mu_{\bot}} \ln Q^{2}/m^{2} + C^{\mu_{\bot}}]+{\mathcal O}(1/Q^{7}). 
\label{ABC}
\ee   
where the leading power $Q^{-6}$ is obtained from the  dimension reasons.
In the absence of the divergent soft contribution we could expect that  amplitudes  $A$ and $B$ can be interpreted in terms of convolutions of the LO and NLO hard coefficient functions with the two- and one-loop evolution kernels, respectively. However, the  divergency of the soft convolution integral in  (\ref{F1s:int}) allows us to suggest that  there is a large logarithm of $Q$ which has a different interpretation. 

The exact calculation of the coefficients $A,\, B,\, C$ in Eq.(\ref{ABC})  is a difficult task involving 2-loop massive integrals and we are not going to proceed in this way.  
We will focus our attention on the possible overlap of the soft and collinear contributions which  can most probably provide the solution of the problem.  
 The idea is to use  the  strategy of regions \cite{Smirnov,Smirnov:2002pj} technique which allows one to find and  interpret 
  the contributions originated from the different sectors in  (\ref{D}). 
  This formalism also allows one  to establish  wether the collinear contributions  overlap with the soft sectors.   

Let us  begin our discussion from the  soft region where 
\be
k_{1\mu}\sim k_{2\mu}\sim\Lambda, 
\label{soft}
\ee
 Corresponding  contribution can be represented as  \cite{Kivel:2010ns}: 
\begin{align}
{D}^{\mu_{\bot}}_{s}  & =  \mathcal{C}\,
\bar\chi_{\bs \beta}\,  \left[ (  \gamma_{\bot}^{\mu}~) _{\beta_{3}\alpha_{3}}\right] _{C_{A}} \chi_{\bs \alpha}
\int dk_{1,2}^{+}
\left[  \frac{~\left[  \gamma^{i}\right]  _{\beta_{1}\alpha'_{1}}\left[ \gamma^{i}\right]
_{\beta_{2}\alpha'_{2}} ~}
{y_{1}\bar{y}_{3}^{2}[-Q(k_{1}^{+}+k_{2}^{+})]^{2}\left[  -Qk_{1}^{+}\right]~}\right] _{J'}
\nonumber\\
&
\int dk_{1,2}^{-}~~
 \left[  \frac{\ ~\left[  \ \gamma^{j}\right]
_{\beta'_{1} \alpha_{1}}~\left[  \gamma^{j}\right]  _{\beta'_{2} \alpha_{2} }  }{~x_{1}\bar{x}_{3}^{2}~[-Q(k_{1}^{-}+k_{2}^{-})]^{2}\left[  -Qk_{1}^{-}\right]  }\right]_{J}  
 \ \left[\int
dk_{12\bot}\frac{\left(  \hat{k}_{1}+m\right)  _{\alpha'_{1}\beta'_{1}}~\left(\hat
{k}_{2}+m \right)_{\alpha'_{2}\beta'_{2}}}{\left[  k_{1}^{2}-m^{2}\right]  \left[
k_{2}^{2}-m^{2}\right]  }\right]_{S}  .
\label{Ds}%
\end{align}
The subscripts $C_{A}, J, S$ are associated with the appropriate contributions in Eq.(\ref{F1s:int}) so that expressions in the brackets  can be associated with  these quantities in pQCD. 

Computing the  transverse integrals inside the brackets associated with SCF $S$ one obtains the answer similar to one  given in Eq.(\ref{SLO}) which is proportional to the mass 
 $m^{2}$.  Redefining the light-cone variables $k_{1,2}^{\pm}$ in the convolution integrals one can eliminate the mass $m$  from the consideration.  
But such redefinition performed  in the divergent integrals may lead to a mistake.  We will see that calculation of the regularized soft integrals  leads to a 
logarithmic dependence on the mass $m$.  

\begin{figure}[th]
\begin{center}
\includegraphics[
height=0.7in
]{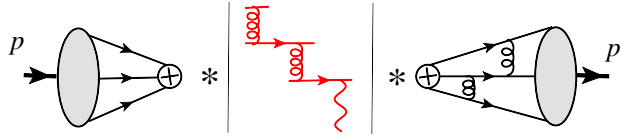}
\end{center}
\caption{Graphical interpretation of the $D^{\mu_{\bot}}_{cp^{\prime}} $  given by  Eq. (\ref{Dcpp}). The hard subdiagram with the two gluon exchange  is separated by vertical line and shown by red lines in the color picture. The asterisk denotes the integral convolution wit respect to collinear fractions.  }
\label{collPp-diagram}
\end{figure}

We find that the contributions of the remaining regions in $D^{\mu}$ can be associated only with the hard rescattering term $F_{1}^{(h)}$ (\ref{F1:fact}).  Corresponding contributions are described  by the different combinations of the collinear and hard regions.  
For brevity we skip the discussion of the all dominant regions and  pass directly to those  which are relevant to our consideration.  These regions  can be described as collinear regions, where both spectator momenta $k_{1,2}$ collinear either to initial $p$ or to final $p'$ momentum.  Such scenario  looks  understandable taking into account that we are looking for overlap with two soft spectators.  
Because of the symmetry between {\it in} and {\it out} states  in Eq.(\ref{D}) it is enough to consider only one of them.  Let us choose
\be
 k_{1}\sim k_{2}\sim p^{\prime}:~k_{i}^{-}\sim Q, k_{+}\sim \Lambda^{2}/Q, k_{\bot}\sim \Lambda .
 \label{collpp}
 \ee  
Then one obtains that corresponding contribution has the leading  power suppression $Q^{-6}$  as in Eq.(\ref{ABC}) and can be written as a convolution 
of hard $[...]_{\text{H}}$ and collinear $\mathcal{V}$ parts  (see details in Appendix C):
\bea
D^{\mu_{\bot}}_{cp^{\prime}}\simeq \mathcal{C}\, \bar\chi'_{\bs \beta}\,
\int dk_{1}^{-}dk_{2}^{-}\, \mathcal{V}_{\bs \beta\bs \beta'}(k_{i}^{-})
\left[\,
\frac{ \left[  ~\gamma^{j}\right]_{ \beta'_{1}\alpha_{1} } 
\left[  ~\gamma^{j}\right]  _{\beta'_{2}\alpha_{2}} \left[  \gamma_{\perp}^{\mu}\right]  _{ \beta'_{3}\alpha_{3}} }
{ \bar{x}_{3}^{2}x_{1}\left[  -Q\left(  k_{1}^{-}+k_{2}^{-}\right)\right]  ^{2}\left[  -Qk_{1}^{-}\right]  }\, \right]_{\text{H}}   \chi_{\bs \alpha} ,
\label{Dcpp}
\eea
where
\bea
\mathcal{V}(k_{i}^{-})  &  = &
\frac12 \int dk_{1}^{+}dk_{2}^{+}~dk_{1\bot}dk_{2\bot}
\frac{
\left\{ \gamma^{i}\left(  \hat k_{1}+m \right)\right\}  _{\beta_{1}\beta'_{1}}
 }
{
\left[  k_{2}^{2}-m^{2}\right]  \left[  k_{1}^{2}-m^{2}\right]
}
\nonumber \\
&  &
\times \frac{
\left\{  \gamma^{i}
(\hat p^{\prime}-\hat p_{3}^{\prime}-\hat k_{1}+m )  \gamma^{\alpha}(\hat k_{2}+m)\right\}_{\beta_{2}\beta'_{2}}
\left\{ \gamma^{\alpha}
 \left(\hat p^{\prime}-\hat k_{1}-\hat k_{2}+m \right)    \right\}  _{\beta_{3}\beta'_{3}}
 }
{ \left[ \left(  p^{\prime}-k_{1}-k_{2}\right)  ^{2}-m^{2}\right]  
\left[  \left(  p^{\prime}-p_{3}^{\prime} -k_{1}\right)^{2}-m^{2}\right]  
\left(  k_{1}+k_{2}-p^{\prime}+p_{3}^{\prime} \right) ^{2}
\left(  k_{1}-p_{1}^{\prime}\right)  ^{2} }.
\label{Vcoll}
\eea
The interpretation of this contribution is given in Fig.\ref{collPp-diagram}

The collinear part $\mathcal{V}$ can be clearly associated with the two-loop contribution to the evolution kernel of the DA. Recall, that  
light-cone components $k_{\pm}$  scale according to (\ref{collpp}). Computing the  integrals over $k_{1,2}^{+}$  in Eq.(\ref{Vcoll}) we obtain 
that the integration region for the minus components  are restricted 
$0< k_{i}^{-}< Q$. Therefore these variables can be rescaled to the dimensionless quantities which, as a rule,  used  as collinear fractions.  
The  integrals over the transverse momenta in  (\ref{Vcoll}) are UV-divergent and these are the logarithmical divergencies  associated with the evolution of DA.

In order to see the overlap of the soft and collinear regions we consider the soft limit (\ref{soft}) in the collinear contribution $D^{\mu_{\bot}}_{cp^{\prime}}$ in Eq.(\ref{Dcpp}). 
In this case  one finds that the collinear-soft limit  reproduces the soft contribution in Eq.(\ref{Ds}):  $(D^{\mu_{\bot}}_{cp^{\prime}})_{s}=D^{\mu_{\bot}}_{s}$.
 This allows us to conclude that collinear and soft regions in this case  overlap. 
 As a consequence,  this also allows us to expect that the convolution integrals with respect to $k_{-}$ in (\ref{Dcpp}) are singular at the end-point region.  
 We will see this explicitly  computing  the collinear integral (\ref{Dcpp}) in Sec.\ref{secIIIB}. 

 Let us emphasize  one important point. Analysis of the soft contribution in Eq.(\ref{Ds})  shows that the  terms
   with momenta $\hat k_{1,2}$ in the numerator of  the  integrand  vanishes so that 
\be
D^{\mu_{\bot}} _{\text{s}}= \mathcal{C}\,
\bar\chi_{\bs \beta}\, \left[ (  \gamma_{\bot}^{\mu}~) _{\beta_{3}\alpha_{3}}\right] _{C_{A}}  \chi_{\bs \alpha}
\int dk_{1,2}^{+}
\left[  \cdots  \right] _{J'}
\int dk_{1,2}^{-} \left[ \cdots \right]_{J}   
\left[ \int~dk_{1\bot}dk_{2\bot}~
\frac{~m^{2}\left( 1 \right)  _{\beta_{1}\alpha_{1}}~(1)_{\beta_{2}\alpha_{2}}}
{\left[  k_{2}^{2}-m^{2}\right]  \left[   k_{1}^{2}-m^{2}\right]  }\right]_{S},
\ee
This can be easily seen if one uses the Sudakov decomposition for the  soft momenta $k_{1,2}$ in the soft part in Eq.(\ref{Ds}). 
Then the terms with the  longitudinal components $k_{i}^{\pm}\gamma_{\mp}$ vanish due  Dirac projections on the leading twist DAs (\ref{LTP}).  
The part with the transverse projections $\Dslash k_{1\bot} \Dslash k_{2\bot}$ vanishes due to the rotation invariance
 in transverse subspace (see also Appendix B). 
 Therefore we obtain that only $\sim m^{2}$  part can provide nontrivial contribution. 
The specific role  of the mass squared term in  discussion of the soft spectator 
scattering was also noted in  \cite{Dun1980}.
  
From this observation we  can  conclude  that the appropriate part  of the collinear contribution   
 $D^{\mu_{\bot}}_{cp^{\prime}}$ in Eq.(\ref{Dcpp}) describing  the soft collinear overlap is UV-finite.  
 Thus  we obtain  that the intersection of the soft collinear regions  in the given example  does not involve
 to the UV-divergencies  of the  collinear operator  which is associated with  the QCD evolution of nucleon DA.

Taking into account  this fact we can conclude that the convolution integrals of the 
  hard coefficient function {\bf H} in the factorization  formula (\ref{F1:fact}) with 1- and 2-loop evolution kernels $\mathcal{ V}_{1,2}$  
  or with the models of  DA  with the vanishing end-point behavior (\ref{end-p}) (for instance, like  (\ref{phi})) are free from the end-point singularities.  
  But the convolution of LO hard coefficient function $H$ in Eq.(\ref{Dcpp})  with the  finite $m^{2}$-term  originating from the collinear subdiagram
   $\mathcal V$ is singular.  Below  we will demonstrate that this singularity provides a  logarithmic contribution which  
  in the sum  with the UV-logarithm from the soft term (\ref{Ds}) yields  a contribution 
  proportional  to $\ln Q/m$.

 Such situation turns out to be different  from the many well known cases when the  end-point singularities appear  after the inspection of the 
 end-point behavior of the evolution kernel, or equivalently, asymptotic DA.  Such situation takes place, for instance, in the case of Pauli FF $F_{2}$. 
 Usually in these cases one deals with  the higher twist DAs.  These DAs describe the higher (excited) components of the hadronic wave function and, 
 as a rule, can be associated with the orbital motion  of the partons. 
  
 But in case of $F_{1}$  the situation is different.  We always consider only the leading twist projections  
 and  therefore we do not involve any  higher twist  DAs. 
 Therefore the mechanism of the factorization breaking in this case  is different.   
 Taking into account the role of  the soft quark masses  we can conclude that in this case  violation 
 of the collinear factorization can be associated with the  chiral symmetry breaking.  
 Therefore we can conclude that   combination of two possibilities: soft spectator scattering together with  
 chiral symmetry breaking provides  the specific  mechanism in pQCD which  leads to the  soft   
 collinear overlap  and breaks the collinear factorization.  This scenario also explains why we do not see
 any problem with the collinear convolution integrals when we use DAs associated with the evolution kernels.
 The QCD evolution is not sensitive  to the chiral symmetry breaking: the evolution kernels can be computed  
 with the massless quarks.   But  how  the chiral symmetry breaking  can affect the DA  in order to provide  
 the singular convolution?  In order to answer this question we have to  find the interpretation of the 
  UV-finite contribution  arising  in $\mathcal{V}$ in Eq.(\ref{Dcpp}) .
  
 Let us consider  the  perturbative analog of the proton DA, i.e. as we would substitute three collinear quark states instead of protons. 
  In such case the 3-quark DA is defined as pQCD matrix element and its  perturbative expansion schematically 
  read\footnote{For brevity we skip the Dirac and color indices }:
\be
\bs \Psi_{\text{PT}}(x_{i},\mu_{F})=\Psi_{0}(x_{i})+\frac{\alpha_{s}(\mu_{R})}{\pi}~\Psi_{1}(x_{i},\mu
_{F})+\left(  \frac{\alpha_{s}(\mu_{R})}{\pi}\right)  ^{2}~\Psi_{2}(x_{i},\mu_{F})+~...~.
\ee
where the higher order coefficients $\Psi_{i}(x_{i},\mu_{F})$, $i>0$  depend logarithmically on the renormalization (or factorization) scale $\mu_{F}$. 
Schematically this can be represented as:
\bea
\Psi_{1}(x_{i},\mu_{F})&=&\ln\mu_{F}^{2}/m^{2}~\mathcal{V}_{1}\ast\Psi_{0}+\Psi
_{10}(x_{i}),~\
\\
\Psi_{2}(x_{i},\mu_{F})&=& \ln^{2}\mu_{F}^{2}/m^{2}~\mathcal{V}_{1}\ast
\mathcal{V}_{1}\ast\Psi_{0}+\ln\mu_{F}^{2}/m^{2}~\mathcal{V}_{2}\ast\Psi_{0}+\Psi_{20}(x_{i}) +\dots ,
\label{Psi2}
\eea
where asterisk as usually denotes  the collinear convolution integrals, $\mathcal{V}_{1,2}$ denote now 1- or 2-loop  evolution 
kernels and dots represent the other contributions associated with the renormalization of QCD coupling.  If we put
$\mu_{F}=m$ then all large logarithms vanish and we obtain:
\be
\bs \Psi_{\text{PT}}(x_{i},m)=\Psi_{0}(x_{i})+\frac{\alpha_{s}(m)}{\pi}~\Psi_{10}(x_{i})+\left( \frac{\alpha_{s}(m)}{\pi} \right)^{2}~\Psi_{20}(x_{i})+~...~.~
\label{init}
\ee
Obviously, Eq.(\ref{init})  is  understood as a perturbative expansion of the DA at the low energy   scale $m\ll Q$.  
Performing transition from perturbative consideration to physical FF  one substitutes instead of perturbative DA the realistic one: 
$\bs \Psi_{\text{PT}}(x_{i},m)\rightarrow \bs \Psi(x_{i},\mu_{0})$,  where it is natural to assume that $\mu_{0}$ is a certain soft scale of order $\Lambda$.  
Following these arguments we can   perform the interpretation of the UV-finite term $\sim m^{2}$ in Eq.(\ref{Vcoll}). 
This term provides the contribution  to  $\Psi_{20}(x_{i})$ and therefore can be associated  with the  DA at  low normalization $\mu_{0}$. 
If   $\Psi_{20}(x_{i})$  decreases  more slowly  at small values  $x_{i}\rightarrow 0$  comparing to  the asymptotic  DA  $\bs \Psi_{as}(x_{i}\rightarrow 0)\sim x_{i}x_{j}$ 
  then the corresponding  collinear convolution integrals  can be singular.  Below we will demonstrate that in pQCD  one obtains the linear behavior  
  \be
  \Psi_{20}(x_{i}\rightarrow 0)\sim x_{i},
  \ee
  assuming that  the ratio of the small fractions $x_{i}/x_{j}$ is fixed.  
  Let us emphasize again  that  such a behavior  is strongly motivated by the presence of  soft spectator contribution.   

  Concluding let us mention that  similar analysis can be also carried out  for  other QCD diagrams which have both  contributions associated with the  hard and soft spectator scattering.  We  are not going  to present it  for all of them because  it does not provide us with any new principal information, on the other hand 
 such analysis is very lengthy and  technically complicate.   But in the next section  we will continue to study  the overlap of the collinear and soft sectors  
 in the framework of pQCD   and present the explicit results for the  collinear and soft contributions discussed above.

\subsection{ Calculation of the large logarithmic term arising from the overlap of collinear and soft regions }
\label{secIIIB} 

Having established that for the soft-collinear overlap one faces  the problem separating unambiguously  the regions in the factorization formulas (\ref{F1:fact}) and (\ref{F1s:int}).   
This is quite  a  complicated problem, especially for the  processes involving   composite particles like hadrons. Therefore it is useful as a first step  to consider more simple 
examples with similar features.  We will continue the  discussion of the diagram in Fig.\ref{2-loop-D}  and  discuss  the separation of the soft and collinear contributions  in this particular case.

 In order to compute the contributions of the soft (\ref{Ds}) and collinear (\ref{Dcpp}) regions explicitly we need a 
specific regularization for the convolution  integrals over  longitudinal momenta. Dimensional regularization (DR)  can not be used in this situation because the soft contribution  is given 
by scaleless integral and therefore equals to zero in this case.  Therefore  one has to introduce  a different regularization in order to work with the collinear and soft integrals in 
(\ref{F1:fact}) and (\ref{F1s:int}).  
This is not only a technical problem. A careful prescription is required in order to avoid double counting computing  the different contributions.     
These  questions  have been studied during last years in many publications in connection with the Sudakov form factor, see e.g. \cite{Collins:1999dz, Idilbi:2007ff, Chiu:2009yx}.  
In \cite{Collins:1999dz} a systematic subtraction procedure has been  suggested  in order to separate contributions of collinear and soft modes in non-inclusive processes.    
The factorization  in the context of  SCET  was discussed in  \cite{Manohar:2006nz}  where the idea of the so-called  zero-bin subtractions
  was invented.   In our analysis we adopt this technique  in order to formulate a  prescription for the correct separation of the soft and collinear modes and compute required integrals. 
  
Note  that  the soft  contribution in Eq.(\ref{Ds}) and the appropriate part of  the collinear  term in Eq.(\ref{Dcpp})   do not overlap with the hard region. As a result  these integrals can be considered as UV-finite (in a sense that  integrals over transverse momenta are finite).  This  allows us to carry out all calculations using a specific regularization in four dimensions which may be the simplest solution in this case.  Such regularization prescription must be formulated uniformly for the all  collinear and soft divergent integrals.  There is also a technical problem about   applicability of the method of regions  with this specific regularization in $D=4$.  In order to fix these details  we suggest to investigate a 
simple one-loop integral which  is close to our  situation  and can be easily computed.   
   
\subsubsection{ Collinear and soft contributions  in $D=4$:  one-loop case }
    
     As example consider following   integral:
\be
J =\int dk~\frac{m^{2}}{\left[  k^{2}-m^{2}\right] ^{2}}
\frac{1}{[k^{2}-2(pk)]}\frac{1}{[k^{2}-2(p^{\prime}k)]}=\frac{i\pi}{Q^{2}}\ln Q^{2}/m^{2}+O(m^{2}/Q^{2}).
\label{J}
\ee
where we assume that expressions in the square brackets are defined with the $+i\varepsilon$ prescription: $[X]\equiv[X+i\varepsilon]$,  and the momenta $p$ and $p'$ are the same as used before. It is easily to see that this integral can be  related to a well-known scalar vertex  integral: 
\be
J= m^{2}\frac{d}{dm^{2}}\int dk~
\frac{1}{\left[  k^{2}-m^{2}\right]  }\frac{1}{[k^{2}-2(pk)]}\frac{1}{[k^{2}-2(p^{\prime}k)]}
\ee
The asymptote of the vertex integral is given by large Sudakov logarithm $\ln^{2} Q^{2}/m^{2}$  but the mass differentiation  reduces this structure  to a simple logarithm.  Analysis of  the dominant regions yields:%
\bea
J_{\text{coll}}\sim J_{s}  \sim\frac{1}{Q^{2}},\quad  J_{h}\sim\frac{m^{2}}{Q^{4}},
\eea
where the subscripts ``coll'', ``s'' and ``h'' denotes collinear to $p$ or $p'$, soft and hard regions respectively. The hard region is suppressed and the large logarithm is generated 
only from the overlap of  collinear and soft regions.  In order to obtain  the leading order result (\ref{J}) with the help of the method of regions we must introduce appropriate regularization.
 It is  not difficult to see that DR can not  help in this case.    

Consider the regularization by small off-shell external momenta introducing the small transverse components.  This yields:
\be
J_{\text{reg}}   =\int dk~\frac{m^{2}}{\left[  k^{2}-m^{2}\right] ^{2}}
\frac{1}{[k^{2}-2(pk)-p_{\bot}^{2}]}\frac{1}{[k^{2}-2(p^{\prime}k)-p_{\bot}^{\prime2}]}
\ee
As usually, one may expect that   the exact answer (\ref{J}) can be reproduced by the sum of the  contributions from the dominant regions:
\be
J\simeq J_{cp'}+J_{cp}+J_{s}.
\label{sumJ}
\ee
But in this case this not true and this can be easily seen from the  explicit calculation.  
The contribution from the soft region reads:
\be
J_{s}\simeq \int dk~\frac{m^{2}}{\left[  k^{2}-m^{2}\right]  ^{2}}\frac{1}{\left[
-p_{+} k_{-}-p_{\bot}^{2}\right]  }
\frac{1}{\left[  -p^{\prime}_{-}k_{+}-p_{\bot}^{\prime2}\right]  }
=
\frac{1}{Q^{2}}\int dk~\frac{m^{2}}{\left[  k^{2}-m^{2}\right]^{2}}
\frac{1}{\left[  -k_{-}-\tau_{-}\right]  }\frac{1}{\left[  -k_{+}-\tau_{+}\right]  }%
\label{Js0}
\ee%
where we introduced  $\tau_{-}=p^{2}_{\bot}/p_{+}$ and $\tau_{+}=p^{\prime 2}_{\bot}/p^{\prime}_{-}$.  Notice that  in the absence of the regulators this soft integral is scaleless 
and has both UV and IR divergencies as the integral in Eq.(\ref{JS}).  Performing integration over $k_{-}$ by residues and taking the transverse 
integral one obtains 
\be
J_{s}=\frac{i\pi}{Q^{2}}\int_{0}^{\infty}d k_{+}~\frac{1}{\left[ k_{+}+\tau_{+}\right]
}\frac{1}{\left[  1+k_{+}\tau_{-}/m^{2}\right]  }=
-\frac{i\pi  }{Q^{2}}\frac{m^{2} \ln\left[  \tau_{+}\tau_{-}/m^{2}\right]  }{\tau_{+}\tau_{-}-m^{2}}\simeq -\frac{i\pi}{Q^{2}}\ln\left[ \tau_{+}\tau_{-}/m^{2}\right] .
\label{Js}
\ee
From Eq.(\ref{Js}) it is clearly seen that the regulators $\tau_{\pm}$  serves as IR- and UV-regulators ($k_{+}\rightarrow 0$ or $k_{+}\rightarrow \infty$ respectively).  Passing to the last equation in (\ref{Js}) we neglected the regular in $\tau_{+}\tau_{-}$ contributions assuming  $\tau_{+}\tau_{-}\ll m^{2}$.

In collinear region $k\sim p^{\prime}$ one obtains
\bea
J_{cp^{\prime}}&=&\int dk~\frac{m^{2}}{\left[  k^{2}-m^{2}\right] ^{2}}
\frac{1}{\left[  -2(pk)-p_{\bot}^{2}\right]  }\frac{1}{\left[  (p^{\prime}-k)^{2}-p_{\bot}^{\prime2}\right]  }
\\
&=&\frac{1}{p_{+}} \int dk_{-} \frac{1} { \left[  -k_{-}-\tau_{-} \right]}~\frac12 \int dk_{\bot} dk_{+} \frac{ m^{2}} { \left[  k^{2}-m^{2}\right]^{2}  } \frac{1}{\left[  k^{2} - 2p^{\prime}_{-}k_{+} \right]  }.
\label{Jcpp}
\eea
We neglected the second  regulator $\tau_{+}$ in  (\ref{Jcpp}) because the corresponding integral is finite.  Again, computing  the integrals over $k_{+}$ and $k_{\bot}$  we obtain
\be
J_{cp^{\prime}}=\frac{i\pi}{p_{+}} \int_{0}^{p_{-}^{\prime}} dk_{-} \frac{1} { \left[  k_{-}+\tau_{-} \right]}=-\frac{i\pi}{Q^{2}}\ln\frac{\tau_{-}}{p_{-}^{\prime}}.
\ee 
Similarly one computes the second collinear integral $k\sim p$:
\be
J_{cp}=-\frac{i\pi}{Q^{2}}\ln\frac{\tau_{+}}{p_{+}}.
\label{Jcp}
\ee
However the sum of the all terms (\ref{sumJ}) can not reproduce the exact answer (\ref{J}).  The reason is that  collinear integrals $J_{cp'}$ and $J_{cp}$  obtain contribution also from 
 the soft region. Computing the soft limit in expression in (\ref{Jcpp}) 
\be
\left( J_{cp^{\prime}}\right)_{s} = 
\int dk~\frac{m^{2}}{\left[  k^{2}-m^{2}\right] ^{2}} \frac{1}{\left[  -2(pk)-p_{\bot}^{2}\right]  }\frac{1}{\left[  -2(p^{\prime}k)-p_{\bot}^{\prime2}\right]  } 
\label{Jcpps}
\ee  
 we reproduce expression for the soft integral in (\ref{Js0}). Hence in order to compute the contribution of the collinear regions correctly one has to subtract from the expressions for collinear integrals  $J_{cp'}$ and $J_{cp}$ appropriate soft contributions \cite{Collins:1999dz} (or similarly to perform zero-bin subtractions \cite{Manohar:2006nz}).  With such subtractions one has
\be
[J_{cp'}-J_{s}] =\frac{i\pi}{Q^{2}}\ln \frac{p^{\prime}_{-}\tau_{+}}{ m^{2}}, \quad    [J_{cp}-J_{s}]=\frac{i\pi}{Q^{2}}\ln \frac{p_{+}\tau_{-}}{ m^{2}} ,
\label{Jcs}
\ee  
Notice that  logarithms in (\ref{Jcs}) originates from the UV-region because IR singularities  cancel in the difference exactly   as it was discussed in \cite{Manohar:2006nz}.
Now the sum of the all contributions reproduces the correct answer:
\be
J= [J_{cp'}-J_{s}]+[J_{cp}-J_{s}]+J_{s} =\frac{i\pi}{Q^{2}}\ln \frac{p^{\prime}_{-}\tau_{+}}{ m^{2}}+\frac{i\pi}{Q^{2}}\ln \frac{p_{+}\tau_{-}}{ m^{2}}  
-\frac{i\pi}{Q^{2}} \ln \frac{ \tau_{+}\tau_{-}}{m^{2}}  =\frac{i\pi}{Q^{2}}\ln\frac{Q^{2}}{m^{2}} .
\ee   
One can easily see that  zero-bin subtractions (\ref{Jcpps})  can be also taken into account  by  changing the sign of the soft contribution in Eq.(\ref{sumJ}) that   was noted already in \cite{Manohar:2006nz, Chiu:2009yx}.  
 
 From considered example we can conclude, that evaluations of integrals using the method of regions in $D=4$ with specific regularization, which 
  allows to avoid scaleless integrals, must be carried out  with the proper IR-subtractions.  In the dimensional regularization  such subtractions as a rule are   performed automatically when one neglects  the scaleless integrals, see for detailed discussion  \cite{Jantzen:2011nz}.

\subsubsection{ Collinear and soft contributions in two-loop integral (\ref{D}) }

Now we  return to the calculation of the overlapping soft and collinear contributions for the more complicate two-loop integral (\ref{D}). 
From  discussion in Sec.\ref{secIIIA}  it is  clear that  the situation 
 with the soft-collinear overlap in (\ref{D}) is quite similar to the considered above one-loop example: 
  in the 2-loop case we also expect to obtain simple a logarithm $\ln Q^{2}/m^{2}$ originating from overlap of the collinear and soft 
 regions. This  allows us  to write
 \be
 D^{\mu_{\bot}}_{s}+\text{UV-finite part}\left[ D^{\mu_{\bot}}_{cp'}+D^{\mu_{\bot}}_{cp} \right]=B_{sc}^{\mu_{\bot}} \ln\frac{Q^{2}}{m^{2}},
 \label{sumD}
 \ee
 where  $B_{sc}^{\mu_{\bot}}$ denote  appropriate part of the total coefficient $B^{\mu_{\bot}}$  in (\ref{ABC}). 
 Computing $B_{sc}^{\mu_{\bot}}$ we can  follow the same line as in the one-loop case and perform all calculations in $D=4$. 

 In order to regularize the divergencies we introduce an infinitesimal gluon mass $\mu$.  
 Such regularization looks quite natural in this case because  the gluon mass plays the role of  the virtuality cut-off  for  hard gluons in the hard subdiagram. 
  Then the regularized  soft contribution (\ref{Ds})  can be written as 
 \begin{align}
{D}^{\mu_{\bot}}_{s}  =  \mathcal{C}\, \bar\chi_{\bs \beta}\, [ \gamma_{\bot}^{\mu}] _{\beta_{3}\alpha_{3} } \, \chi_{\bs \alpha} 
& \int dk_{1,2}^{+}
 \frac{~
 \left[  \gamma^{i}\right]  _{\beta_{1}\alpha'_{1}}\left[ \gamma^{i}\right]
_{\beta_{2}\alpha'_{2}}
~}
{\bar{y}_{3}[-Q(k_{1}^{+}+k_{2}^{+})][-{y}_{3} Q(k_{1}^{+}+k_{2}^{+})-\mu^{2}]\left[  -y_{1} Q k_{1}^{+}-\mu^{2}\right]~}
\nonumber\\
\times &
\int dk_{1,2}^{-}~
  \frac{
  ~\left[  \ \gamma^{j}\right]_{\beta'_{1} \alpha_{1}}\left[  \gamma^{j}\right]  _{\beta'_{2} \alpha_{2} } 
  }
{~\bar{x}_{3}~[-Q(k_{1}^{-}+k_{2}^{-})][-{x}_{3}Q(k_{1}^{-}+k_{2}^{-})-\mu^{2} ] \left[  -x_{1} Q k_{1}^{-}-\mu^{2} \right]  } 
\nonumber \\
\times & \int
dk_{12\bot}
\frac{
m^{2} \left( 1 \right)  _{\alpha'_{1}\beta'_{1}}\left(1 \right)_{\alpha'_{2}\beta'_{2}}
}
{\left[  k_{1}^{2}-m^{2}\right]  \left[ k_{2}^{2}-m^{2}\right]  } .
\label{Dsreg}%
\end{align}
where we used  that in Breit system $p'_{-}\simeq p_{+}\simeq Q$, see Appendix A.  Calculation of these integrals is a bit tedious but follows a basic line: two integrations performed by residue and the remnant integrals can be further computed keeping the most singular at $\mu\rightarrow 0$ terms.  The details can be found  in Appendix D. The result reads
\bea
{D}^{\mu_{\bot}}_{s} & = & \mathcal{C}\,  \bar\chi_{\bs \beta}\, 
[\gamma^{i}\gamma^{j}]_{\beta_{1}\alpha_{1}}    
[\gamma^{i}\gamma^{j}]_{\beta_{2}\alpha_{2}}
[ \gamma_{\bot}^{\mu}] _{\beta_{3}\alpha_{3}} \, \chi_{\bs \alpha}  \, 
\frac{(2\pi i)^{2}}{Q^{6}}  \frac{1}{x_{1} \bar{x}_{3}^{2}}\frac{1}{y_{1}\bar{y}_{3}^{2}}
 \left(  1-\frac{\pi^{2}}{6} \right)  \ln\frac{  \tau_{+}\tau_{-} }{m^{2}}   +\mathcal{O}(1),
\label{Ds:res}
\eea
where we again  used  $\tau_{+}=\mu^{2}/p'_{-}$ and $\tau_{-}=\mu^{2}/p_{+}$.

Consider now the collinear term (\ref{Dcpp}). We pick up from the collinear kernel  $\mathcal V$ only UV-finite  part  (UV-f.p.) relevant for our calculation. Then the collinear integral reads
\bea
\text{UV-f.p. } D^{\mu_{\bot}}_{cp^{\prime}}\simeq   \mathcal{C}  \, 
 \int  dk_{1}^{-}dk_{2}^{-}
\frac{ { \Psi}_{20}^{\bs \beta}(k_{i}^{-})\,  [\gamma^{j}]_{\beta_{1}\alpha_{1}}    
[\gamma^{j}]_{\beta_{2}\alpha_{2}}
[ \gamma_{\bot}^{\mu}] _{\beta_{3}\alpha_{3}} \, \chi_{\bs \alpha}  }
{ \left[  -\bar{x}_{3} Q\left(  k_{1}^{-}+k_{2}^{-}\right) \right]  \left[  -\bar{x}_{3}Q\left(  k_{1}^{-}+k_{2}^{-}\right)-\mu^{2} \right]  \left[  -x_{1}Qk_{1}^{-}-\mu^{2} \right]  }
 ,
\label{Dcppreg}
\eea
with 
\bea
{ \Psi}^{\bs \beta}_{20}(k_{i}^{-})  &  = &
\bar \chi'_{\beta'_1\beta'_2\beta_3}[ \gamma^{i} ]_{\beta'_{1}\beta_{1}}
[ \gamma^{i} ]_{\beta'_{2}\beta_{2}}
\int dk_{1}^{+}dk_{2}^{+}~dk_{1\bot }dk_{2\bot}
\frac{ m^{2}  }
{  \left[  k_{2}^{2}-m^{2}\right]  \left[  k_{1}^{2}-m^{2}\right]  }
\nonumber \\ &  &
\times\frac{ -2(p' k_{1})}
{ \left[ \left(  p^{\prime}-k_{1}-k_{2}\right)  ^{2}-m^{2}\right]  
\left[  \left(  p^{\prime}-p_{3}^{\prime} -k_{1}\right)^{2}-m^{2}\right]  
\left(  k_{1}+k_{2}-p^{\prime}+p_{3}^{\prime} \right) ^{2}
\left(  k_{1}-p_{1}^{\prime}\right)^{2} }.
\label{Psi20}
\eea
The notation $\Psi_{20}$ is introduced taking into account  the structure of  DA  described in Eq.(\ref{Psi2}).     
Recall that according to (\ref{Dcpp} ) the integrals in Eq.(\ref{Dcppreg}) represent  the convolution of the hard coefficient function with the perturbative DA  $\Psi_{20}$.  
We carried out the calculations of the integrals in Eq.(\ref{Psi20}) keeping only the most singular contributions  at the limit $k_{i}^{-}\sim 0$. The details are described  in Appendix E. 
The result can be written as
\bea
{ \Psi}^{\bs \beta}_{20}(k_{i}^{-}) &=& 
\bar \chi'_{\beta'_1\beta'_2\beta_3}[ \gamma^{i} ]_{\beta'_{1}\beta_{1}}
[ \gamma^{i} ]_{\beta'_{2}\beta_{2}}
 \frac{(-1)(2\pi i)^{2} }{y_{1}\bar{y}_{3}^{2}Q^{3}} 
\nonumber  \\ & & \times
   \theta(0<k^{-}_{1}/Q<  k^{-}_{1max} )\theta(0<k^{-}_{2}/Q<k^{-}_{2max} )
k^{-}_{2}\ln\left(1+\frac{k^{-}_{1}} {k^{-}_{2}}\right)+\dots,
\label{Psi20:res}
\eea
where dots denote the regular at the limit $k_{i}^{-}\sim 0$ contributions which we do not consider for simplicity; the quantities
  $k^{-}_{1,2max}\sim 1$ denote  the upper boundary for the relative light-cone components $k_{i}^{-}/p^{\prime}_{-}$. Their explicit value is irrelevant because we compute the integral in Eq.(\ref{Dcppreg}) 
with the leading logarithmic accuracy. 
Notice that  from the expression (\ref{Psi20:res}) one can see that  the end-point behavior of ${ \Psi}_{20}$ at small fractions  vanishes {\it linearly}  (  $k_{i}^{-}\rightarrow 0$ and ratio $k^{-}_{1}/k^{-}_{2}$  is fixed )
 \bea
 { \Psi}_{20}(k_{i}^{-} \rightarrow 0) \sim  k^{-}_{2}.
 \eea
 This behavior differs from the  quadratical  asymptote $\sim  k^{-}_{1}k^{-}_{2}$ expected from the  QCD evolution.  
 As a result the collinear convolution integral is singular as discussed in Sec.IIIA. 

In order to see this  we  substitute (\ref{Psi20:res}) into (\ref{Dcppreg}) and compute the regularized convolution integrals. For simplicity,  we shall  take into account the appropriate zero-bin subtractions  changing the sign of  soft contribution (\ref{Ds:res}) in the sum (\ref{sumD}). 
Then we have
\bea
\text{UV-f.p.} {D}^{\mu_{\bot}}_{cp'}  =  \mathcal{C} 
 \bar\chi_{\bs \beta}\, 
[\gamma^{i}\gamma^{j}]_{\beta_{1}\alpha_{1}}    
[\gamma^{i}\gamma^{j}]_{\beta_{2}\alpha_{2}}
[ \gamma_{\bot}^{\mu}] _{\beta_{3}\alpha_{3}} \, \chi_{\bs \alpha}  \,  
 J_{cp'},
\eea
where 
\begin{align}
J_{cp^{\prime}}  &  = 
\frac{(2\pi i)^{2}}{Q^{6}}
\frac{1}{x_{1}\bar{x}_{3}^{2}}     \frac{1}{y_{1}\bar{y}_{3}^{2}}
\int_{0}^{k^{-}_{1max}}dk^{-}_{1}     \int _{0}^{k^{-}_{2max}} dk^{-}_{2}
\frac{k^{-}_{2}\ln(1+k^{-}_{1}/k^{-}_{2})}{~\left(  k^{-}_{1}+k^{-}_{2}\right)  ^{2}\left[  k^{-}_{1}+\tau_{-}/x_{1}p'_{-}\right]  }
\\
&  \simeq \frac{(2\pi i)^{2}}{Q^{6}} 
\frac{1}{x_{1} \bar{x}_{3}^{2}}  \frac{1}{y_{1}\bar{y}_{3}^{2}}
\int_{0}^{k^{-}_{1max}} \frac{dk^{-}_{1}}{k^{-}_{1}+\tau_{-}/x_{1}p'_{-} }
\int_{0}^{\infty}dk^{-}_{2}\frac{k^{-}_{2}\ln(1+1/k^{-}_{2})}{~\left(1+k^{-}_{2}\right)  ^{2}}
 \\
& \simeq
\frac{(2\pi i)^{2}}{Q^{6}}  \frac{1}{x_{1} \bar{x}_{3}^{2}}\frac{1}{y_{1}\bar{y}_{3}^{2}}
\left(  1-\frac{\pi^{2}}{6}\right)  \ln \tau_{-}/p'_{-} ~.
\label{Jcp:res}
\end{align}
 The similar calculation for the second collinear integral yields
\be
\text{UV-f.p.} {D}^{\mu_{\bot}}_{cp}  =  \mathcal{C} 
 \bar\chi_{\bs \beta}\, 
[\gamma^{i}\gamma^{j}]_{\beta_{1}\alpha_{1}}    
[\gamma^{i}\gamma^{j}]_{\beta_{2}\alpha_{2}}
[ \gamma_{\bot}^{\mu}] _{\beta_{3}\alpha_{3}} \, \chi_{\bs \alpha}  \, \frac{(2\pi i)^{2}}{Q^{6}}  \frac{1}{x_{1} \bar{x}_{3}^{2}}\frac{1}{y_{1}\bar{y}_{3}^{2}}
\left(  1-\frac{\pi^{2}}{6}\right)  \ln \tau_{+}/p_{+} .
\ee  

Substituting obtained results into (\ref{sumD}) and changing sign in front of the soft term we obtain
\bea
B_{sc}^{\mu_{\bot}}\ln\frac{Q^{2}}{m^{2}} = \mathcal{C} 
 \bar\chi_{\bs \beta}\, 
[\gamma^{i}\gamma^{j}]_{\beta_{1}\alpha_{1}}    
[\gamma^{i}\gamma^{j}]_{\beta_{2}\alpha_{2}}
[ \gamma_{\bot}^{\mu}] _{\beta_{3}\alpha_{3}} \, \chi_{\bs \alpha} \,
 \frac{(2\pi i)^{2}}{Q^{6}}  \frac{1}{x_{1} \bar{x}_{3}^{2}}\frac{1}{y_{1}\bar{y}_{3}^{2}}\left(  1-\frac{\pi^{2}}{6}\right)  \ln\frac{m^{2}}{Q^{2}}.
 \label{Bsc:res}
\eea
We see that all $\tau$-regulators cancel as they should and we obtain a simple large  logarithm.  Recall that the collinear 
contributions are associated with the hard rescattering
term $F_{1}^{(h)}$.   Therefore this calculation explicitly demonstrates that  the soft and hard  spectator scattering contributions are related 
and must be computed consistently.  

To summarize this section, we  demonstrated that the 2-loop diagram with massive quarks  has a large logarithmic term originating from the  overlap of the soft and collinear regions.  The appearance of this logarithm  does not contradict to the discussed  factorization scheme.  
We demonstrated  that in the perturbation theory one can perform the consistent description of large-$Q$ asymptotic of two-loop diagrams using  
definitions of  SCF (\ref{Sdef}) and  DA (\ref{DA:def})  with  collinear quarks instead of nucleon state.  We have also  seen that  collinear integrals  in the hard spectator contribution must be singular and such situation can be realized only due to to the specific  end-point behavior of the perturbative DA. 

From this lesson we can suggest  a general solution of the problem with the end-point divergencies in the soft spectator scattering contribution obtained in Sec.II~ in the perturbation theory.  The  soft spectator scattering together with  chiral symmetry breaking by quark mass $m$ in  QCD  make possible the soft collinear overlap   and as a result the violation of the collinear factorization for the FF $F_{1}$.   
In this case the perturbative DA $\bs \Psi (x_{i},\mu)$ at  low normalization point $\mu=m$  has a such  end-point behavior  which leads to the end-point singularities in the collinear convolution integrals.  At the same time the convolution integral of the hard kernel with the collinear one  remains  well defined.  Such mechanism is not suppressed by inverse power of $Q$ because the quark mass is relevant only for in definition of  the soft correlation function.  

 This allows us to expect that the  factorization of the realistic hard  $F_{1}^{(h)}$ and soft  $F_{1}^{(s)}$ contributions in the requires an additional regularization for the separation of collinear and soft  sectors.  
Such separation has been carried out for the case of 2-loop diagram but a  realization  of similar scheme for physical form factor  involving the different nonperturbative matrix elements  requires farther work.

\section{Phenomenological  applications of  SCET factorization }

From the consideration of Secs.~II~and~III we can conclude that the practical realization of the factorization scheme discussed in the introduction has some difficulties.  If we assume that the nucleon DA at low normalization point has specific non-asymptotic end-point behavior as discussed in Sec.III   then the collinear and soft convolution integrals are not defined in the both equations (\ref{F1:fact})  and  (\ref{F1s:int}).  
Such situation in a some sense is similar to the  semi-inclusive  factorization with the transverse-momentum dependent PDFs, see e.g. \cite{Collins} :  at a formal level  we have established all the dominant regions and provided the formal definitions for the all nonperturbative quantities but that is not enough.  
The complication  arises due to the overlap of the collinear and soft regions  and, as a result,  appearance of the large  rapidity logarithms $\sim \ln Q/\Lambda$.  
  This is well known problem, for many processes where collinear factorization is broken due to singular end-point behavior of the convolution integrals see, e.g., 
\cite{  Manohar:2006nz, Chernyak:1983ej, Bell:2005gw, Lange:2003pk, Beneke:2002bs}. 
The  same situation is also relevant for many other hadronic reactions see, for instance,  recent  discussions  in \cite{Becher:2010tm, Becher:2011pf, GarciaEchevarria:2011rb, Chiu:2012ir}  and references there in.   The specific of  the FF factorization is that the soft functions defined in SCET involve not only  Wilson lines but also quark fields, appearing as a soft spectators.  Therefore there are  two possible configurations  $F_{1}^{(s)}$  and hard $F_{1}^{(h)}$  associated with the different  hard subprocesses.  

 Usually the presence of  two hard scales: the hard $\sim Q^{2}$  and hard-collinear $\sim \Lambda Q$ in the description of nucleon 
 FFs allows one to perform the factorization in two steps. First,  one integrates over the hard  fluctuations and passes from  QCD to SCET-I. 
This effective theory   includes hard-collinear modes which 
can be further  factorized if  the virtualities of the hard-collinear particles are quite large.  Integrating over hard-collinear 
modes one passes from SCET-I to SCET-II  which includes only collinear and soft particles with the virtualities of order $\Lambda^{2}$.  
 However if the value of $Q$ is moderate, (for instance, the hard-collinear scale $Q\Lambda\sim m_{N}\sim 1\text{GeV}^{2}$ is not  large in order to serve as expansion parameter) then one can not  perform SCET-II factorization. Phenomenologically such situation is relevant for quite a large range of experimentally accessible values of momentum transfer. For instance, if  $Q^{2}$ is in the range of $4-25$GeV$^{2}$ one obtains that $Q\Lambda$ varies  between $0.8-2$GeV$^{2}$ assuming  that $\Lambda \simeq 400$MeV.  Taking into account  that  QCD expansion of the jet functions is starting from $\alpha^{2}_{s}$  we can expect that one needs much larger hard-collinear scale in order to see the dominance of this leading contribution.  On the other hand we can not ignore  that the soft  contribution $F_{1}^{(s)}$  is suppressed  by a Sudakov form factor.  If  this suppression is strong enough then then   $F_{1}^{(s)}$  would be subleading compared to the hard spectator contribution $F_{1}^{(h)}$.  However many phenomenological studies indicate that  in the intermediate region of $Q^{2}$ such suppression is still  weak and the soft mechanism provides  essential contribution to the physical FFs.  May be heuristically  this  can  be explained  also by the relative smallness of $F_{1}^{(h)}$ which already at leading order is suppressed as $\alpha^{2}_{s}(Q^{2})$.  Therefore in order to estimate the relative importance of the two terms  it is necessary to include both of them consistently  within SCET-I factorization scheme.  But the end-point singularities in the hard scattering contribution make such program  complicates such program.  Such situation is relevant also for $F_{2}$  and  for many other hard exclusive reactions involving  nucleons. 
In some cases this difficulty can be avoided  using the universality of the SCET-I matrix elements.  
 
 Suppose that we have different scattering  processes  which are described within  SCET-I factorization and depend on the same SCET-I matrix element. Using universality  of  the SCET-I amplitude one can define the so-called {\it physical subtraction scheme}  \cite{Beneke:2000ry, Beneke:2000wa}   which allows one to perform the  systematic  calculations of the hard spectator scattering contributions  associated with the symmetry breaking corrections.  The idea of this approach is very simple: the SCET-I soft-overlap  form factor or amplitude can be rewritten as a sum of one of the physical amplitudes and the corresponding hard spectator contribution. Then this combination can be used further for the analysis of physical amplitudes of other processes with the same SCET-I  matrix elements. After such redefinition the end-point singularities in the combination of the hard spectator terms must cancel and one obtains the well defined hard correction.  This scheme has been successfully used for analysis of different  B-meson decay amplitudes and  we expect that it can  also be used for the analysis of the different hadronic reactions  with the soft spectator scattering contributions. 
 
Let us illustrate the above discussion by one concrete example.   Consider the following processes: $\gamma^{*}N\rightarrow N$ describing proton and neutron form factors at large $Q^{2}$ and   $\gamma^{*}p\rightarrow \pi^{0} p$ describing wide-angle hard electroproduction of pion with $s,t,Q^{2}\gg \Lambda^{2}$.  We suppose that nucleon form factors are described by following tentative formulae:
\bea
F^{p}_{1}(Q)=C_{A}(Q)~\{\, e_{u}~f_{1}^{\text{u}}(Q)+e_{d}~f_{1}^{\text{d}}(Q) \}+\mathbf{\Psi}*\mathbf{H}_{p}*\mathbf{\Psi},
\label{Fp}
\\
F^{n}_{1}(Q)=C_{A}(Q)~\{\, e_{u}~f_{1}^{\text{d}}(Q)+e_{d}~f_{1}^{\text{u}}(Q) \}+\mathbf{\Psi}*\mathbf{H}_{n}*\mathbf{\Psi},
\label{Fn}
\eea
where we used definitions (\ref{f1ud}),(\ref{f1q:def})  and  isotopic symmetry, symbols $\mathbf{H}_{p,n}$ denote the hard scattering kernel for the proton and nucleon cases, respectively.  We also assume that  the convolution integrals denoted by asterisk  are regularized using some IR-regulator. Solving these equations with respect to SCET FFs one finds ( $e_{u}=2/3,\, e_{d}=-1/3$)
\bea
f_{1}^{\text{u}}=C^{-1}_{A}
\left\{
2 F^{p}_{1}+F^{n}_{1}-\mathbf{\Psi}*(2\mathbf{H}_{p}+\mathbf{H}_{n})*\mathbf{\Psi}
\right\},
\label{f1uFH}
\\
f_{1}^{\text{d}}=C^{-1}_{A}
\left\{
F^{p}_{1}+2F^{n}_{1}-\mathbf{\Psi}*(\mathbf{H}_{p}+2\mathbf{H}_{n})*\mathbf{\Psi}
\right\},
\label{f1dFH}
\eea

\begin{figure}[th]
\begin{center}
\includegraphics[height=1.2in]{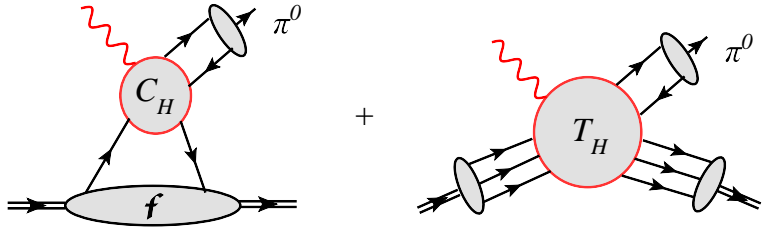}
\end{center}
\caption{The sum of the soft and hard spectator contributions  for the wide-angle hard pion electroproduction}%
\label{gn-pin}%
\end{figure}

The pion production process can also be described  as a sum of two contributions as shown in Fig.\ref{gn-pin}. 
So far we are  not going  to prove  the exact factorization theorem  for this process. Most important for us is that this configuration can be considered as  one possible contribution to the nucleon helicity conserving amplitudes $A^{\pi^{0}}_{i}$.  The first term in Fig.\ref{gn-pin}  describes the soft overlap nucleon contribution and can be expressed in terms of the same SCET-I form factor $f^{u,d}_{1}$. The pion blob is described by pion distribution amplitudes.   The second contribution in Fig.\ref{gn-pin}  can be associated with the hard spectator scattering and it is described as usually as a convolution of nucleon and pion DAs with the hard scattering kernel $T_{H}$.  Therefore  schematic expression of the contribution in Fig.\ref{gn-pin}   reads 
 \bea
A^{\pi^{0}}_{i} =\varphi_{\pi}(z)*C_{H}(z)(e_{u}f_{1}^{\text{u}} -e_{d} f_{1}^{\text{d}})+\mathbf{\Psi}(x_{i})*\varphi_{\pi}(z)*\mathbf{T}_{H}(x_{i},y_{i},z) *\mathbf{\Psi}(y_{i})
\label{Api}
\eea 
where we again assume a some regularization for the divergent convolution integrals. Substitution of expressions for the SCET form factors (\ref{f1uFH}) and (\ref{f1dFH})  in  Eq.(\ref{Api}) yields
\bea
A^{\pi^{0}}_{i} &=&\varphi_{\pi}(z)*C_{H}(z)/C_{A}\frac13( 5F^{p}_{1}+4F^{n}_{1} )
\\
&+&\mathbf{\Psi}(x_{i})* \varphi_{\pi}(z)*\left(\mathbf{T}_{H}(x_{i},y_{i},z)-
C_{H}(z)/C_{A}\frac13\{5\mathbf{H}_{p}(x_{i},y_{i})+4\mathbf{H}_{n}(x_{i},y_{i})\}\right) *\mathbf{\Psi}(y_{i}).
\label{ApiFH}
\eea 
Thus  the soft overlap contribution is represented in terms of the physical FFs $F^{p,n}_{1}$. The  ratio $C_{H}(z)/C_{A}$ depends only from the factorization scale associated with the evolution of  pion DA. All Sudakov logarithms must cancel in this ratio. On the other hand, the end-point singularities in the hard scattering kernels must be compensated in the combination of $\mathbf{T}_{H}$ and $\mathbf{H}_{n,p}$ in rhs of 
(\ref{ApiFH}).  The simple analysis show that the hard spectator correction is of order $\alpha^{3}_{s}$. At the same time the ratio 
$C_{H}(z)/C_{A}=\alpha_{s} C_{\text{LO}}(z)+{\mathcal O}(\alpha^{2}_{s})$.  The next-to-leading contribution can be computed from the one loop corrections to $C_{H}(z)$ and $C_{A}$ and   the hard spectator  corrections can appear only in the next-next-to-leading order. 

Therefore   with this example we demonstrated  how the SCET-I factorization on the soft and hard spectator contributions allows one to analyze the realistic hadronic processes at the intermediate momentum transfer.   Of course, such method can be used when the number of observables is larger then the number of unknown SCET amplitudes.  
 The more detailed  application of this scheme for analysis of different processes  we are going to present in the separate publication.

 If the values of  $Q$ are quite large  so that the hard-collinear scale is already a good parameter for the asymptotic approximation one  can 
 try to perform the second factorization 
 step and pass from the intermediate SCET-I to the low energy  SCET-II.  
 Performing this  matching one must provide a solution of the soft collinear overlap in order to
  treat correctly corresponding large logarithms and avoid a double counting between the soft and hard spectator scattering contributions.

Formulation of such scheme is a difficult task and at present time we have not yet  found a convincing technical realization of this idea which can be realized in our case. 
The most challenging  problem  is the consistent formulation  of  the  zero-bin (or infrared ) subtractions for the collinear convolutions integrals with
 nonpertubative DAs (or other collinear matrix elements) and corresponding UV-renormalization  of the soft rescattering contribution involving soft correlation functions, 
 i.e. unambiguous matching of collinear and soft modes and consequently the definition of the hard and soft rescattering contributions.  
 The other difficult moment is  the absence of a well-defined regularization.  Dimensional regularization can not be used for the calculation of the 
 soft convolution integrals because   corresponding integrands are scaleless.  Therefore  one has to invent  the other method, for instance  off-shell momenta or analytic regularization.  As a rule,  such regularization has potential problems beyond the leading order that makes difficult  systematic calculations and analysis.  
Recently several ideas  has been suggested see, e.g.,  \cite{Becher:2011pf, GarciaEchevarria:2011rb, Chiu:2012ir}    in the framework of SCET approach  which can 
potentially solve these problems. We leave the investigation  of this question  for future work.

\section{Discussion}  

    
 In this paper we  investigated   the large-$Q$ asymptotic form of the proton FF  $F_{1}$ using factorization approach. 
  In our previous publication \cite{Kivel:2010ns} we suggested that the full factorization formula for the nucleon FF $F_{1}$ is given 
 by the sum of two contributions:  hard  spectator scattering  term $F_{1}^{(h)}$  and   the second term describing  the scattering with 
 the soft spectator quarks which we refer as the soft  spectator  contribution $F_{1}^{(s)}$.
 
    In  the  first publication \cite{Kivel:2010ns}  we provided an explicit definition for the leading power 
    contribution $F_{1}^{(s)}$ in terms of the hard and hard-collinear  coefficient functions and  
    nonpertubative matrix elements.  In present  paper we carried out more detailed  analysis of suggested factorization. 
We computed  the leading order jet functions $J$  appearing  in the matching from SCET-I to SCET-II.    
Using this result  we  obtained   that the corresponding soft convolution integrals in definition $F_{1}^{(s)}$  (\ref{F1s:int}) have logarithmic divergency at the end-point region.   
  This observation  follows from the boost invariance  and  therefore does not depend on 
  any model assumptions about  the soft correlation function $\bs S$ in Eq.(\ref{F1s:int}).   

We suggested that the appearance of  end-point singularity  can be explained by overlap of collinear and soft regions.   
 In order to verify this  supposition  we carried out analysis of  two-loop  QCD diagrams with massive quarks  and 
 investigated the dominant contributions associated with the  different regions.  In this case the non-perturbative matrix elements associated with the soft correlation function and distribution amplitudes can be computed explicitly in the perturbation theory.  
 We found  that  the overlap of the soft and collinear sectors  provides the
 large rapidity logarithm  $\sim \ln Q/m$ where $m$ is the quark mass which plays the role of the soft scale.  
 Thus   the  logarithmic divergence of the soft convolution integrals can 
 be naturally explained by the overlap with the collinear region.  
 In such case the collinear convolution integrals describing the hard spectator scattering  has also end-point singularity and can not  be well defined  without  special regularization.  
 The latter  allows us to conclude that  the separation of  the hard and soft
 spectator contributions  for the FF $F_{1}$  is closely  related to the problem of  separation of  soft and collinear sectors in SCET.

 At first glance such situation looks controversial for the FF $F_{1}$  taking into account the end-point behavior of the nucleon DA   which is usually estimated from the QCD evolution.  
The eigenfunctions of the the leading order evolution kernel  vanish quite rapidly at the end-point region  that ensures a good convergence of the collinear integrals defining the hard spectator contribution.
However  the soft  spectator scattering and explicit chiral symmetry breaking (mass of the soft quarks)  in perturbative calculations provides the mechanism which  makes possible 
the soft collinear overlap   and  violation of the collinear factorization without any conflict with these arguments. 
 QCD evolution is not sensitive to the effects related to the mass of quarks   
 and therefore in this case it can not provide a signal about the problem with the end-point  behavior of the collinear convolution integrals. Formally, in the perturbation theory this scenario is realized as following.  The end-point behavior of the DA  at low normalization ($\mu_F=m$)  is different from the one which we expect from its evolution: it vanishes more slowly.   Then the convolution integral with such DA  has logarithmic end-point divergency.       
 
 Extrapolating these arguments  beyond the perturbation theory  we can not use  the mass of the soft quarks as  realistic argument anymore.  But  on the other hand we can expect that  the nontrivial contribution to the soft correlation function $\bs S$ can be obtained due to the dynamical chiral symmetry breaking in non-perturbative  QCD.   In that case  one can have  the nontrivial soft spectator scattering contribution which has  the end-point singularities.   
 Basing on this observation we can  formulate  the  following  conjecture.  
  The  soft spectator scattering and chiral symmetry breaking in non-perturbative QCD  make possible the soft collinear overlap  and as a result violation of the collinear factorization for FF $F_{1}$.   This leads to the appearing of the  end-point singularities in the soft and hard spectator scattering contributions.    
 In this case {\it the nucleon DA $\varphi_{N}(x_{i},\mu_{0})$ at a low normalization point $\mu_{0}\sim \Lambda$  vanishes more slowly  then the asymptotic shape   and such behavior  leads to the end-point singularities in the collinear convolution integrals}. 
Such a behavior is very important for the  consistent description  of the large rapidity logarithm appearing due to overlap of the hard and soft  sectors.   The end-point singularity  arising in the soft spectator contribution can be considered as strong argument in support of such scenario. 
 
We expect that the hypothesis about the role of chiral symmetry breaking in the non-perturbative calculation of the soft correlation function might be verified  in different model calculations. 
 A certain  indication about the non-asymptotic end-point  behavior of the nucleon DA  was already obtained from the non-perturbative calculations.   In work \cite{Petrov:2002jr}  the nucleon DA $\varphi_{N}(z_{i})$ has been computed in the chiral quark-soliton model \cite{Diakonov:1987ty, Diakonov:2000pa} 
  using  large $N_{c}$ approximation.  The DA  is estimated  at $\mu_{0}\leq 600$MeV  and it was found that it  does not vanish at the end-point limit if the one of the fractions, for instance,  $x_{1}$ is fixed and $x_{2}\rightarrow 0$.   Unfortunately the region of small fractions in \cite{Petrov:2002jr}  have not been studied analytically but the authors do not exclude  the linear asymptotic if the all fractions are small $x_{i}\rightarrow 0$ but their ratio is fixed \cite{Petrov:pr}.    On the other hand  in \cite{Petrov:2002jr}   it is also noted  that   obtained results are valid only in the region of relatively large collinear fractions where $x_{i} N_{c}\sim 1$.  It means that  for the realistic value $N_{c}=3$ one can  expect large  $1/N_{c}$  corrections  in  region with small  fractions  and   therefore the obtained 
  leading $1/N_{c}$ results  can be modified.

  Basing on our assumption  we expect that the general scheme as described in Eqs (\ref{F1:fact})-(\ref{F1s:int}) is  valid but  the 
 explicit calculation of the soft $F_{1}^{(s)}$ and hard $F_{1}^{(h)}$  spectator contributions requires  a
 certain prescription for  separation of  soft and collinear degrees of freedom.   
 The collinear convolution integrals in the  $F_{1}^{(h)}$ must be  regularized and
corresponding large rapidity logarithms if  possible must be also resummed.    
  In present paper we do not provide any systematic formalism  
 for  the  factorization of the soft and collinear degrees of freedom. 
 However, we  would like to stress that the soft spectator contribution 
 plays the important role in the correct description of such factorization.  
 
 Our analysis  provides additional arguments  in support of the phenomenological descriptions   which    assumes that at some moderate values of $Q$ 
 the dominant contribution is provided by  soft-overlap contribution  which can be associated with the soft spectator scattering see, {\it e.g.},  
 \cite{Radyushkin:1998rt, Diehl:2002yh, Kroll:2005ni, Feldmann:2011xf}.     
 In spite of  the factorization  complexities the SCET   provides a powerful framework for   investigation    of other hard exclusive reactions.      
 SCET description of the  soft spectator contribution  naturally  introduces a concept of  two large scales: hard $\sim Q^{2}$ and hard-collinear $\sim \Lambda Q$. 
This naturally defines  the region of moderate values of large momentum transfer $Q$:   it corresponds to the situation when the inverse power of hard scale $1/Q^{2}$ 
is a good   expansion parameter but the hard-collinear scale is still not too large. 
  Taking $\Lambda\simeq 300-400$MeV  and $Q^{2}=25$GeV$^{2}$ one easily obtains that 
$Q\Lambda\simeq 1.5-2$GeV$^{2}$. Hence in the situation when $Q^{2}\simeq 10-20$GeV$^{2}$ which overlaps with the 
majority of existed and upcoming experiments,  one can perform  consistently only SCET-I factorization.  
Within such factorization scheme it is useful to take into account both 
soft and hard spectator scattering contributions because  their relative contribution might be comparable if the  suppression 
from the Sudakov form factor  for the soft-overlap term is not sufficiently large.  Using the universality  of the SCET FFs one 
can make  the combined analysis   of the different processes  and  check the relevance of the soft overlap contribution  
 in a model independent way.   Technically it can be done  using  the, so-called, physical subtraction scheme  which allows to 
 to perform a systematic consideration and also solves the problem with the  end-point singularities from the hard spectator corrections. 
In particular,  such analysis  might be very interesting for hard exclusive processes  with baryons, such as wide-angle Compton scattering,  
wide-angle meson production and their  timelike analogs.  This work is in progress.

\section*{ Acknowlegmens}
 This work was supported by the Helmholtz Institute Mainz. 
 The author is grateful to V.~Pascalutsa  and M.~Venderhaeghen for  reading the manuscript.

\section*{Appendix A:  Brief summary of used  notations}

Through the paper  we imply  Breit frame
\begin{equation}
q=p^{\prime}-p=Q\left(  \frac{n}{2}-\frac{\bar{n}}{2}\right)
,~\ n=(1,0,0,-1),~\bar{n}=(1,0,0,1),~\ (n\cdot\bar{n})=2,~
\end{equation}
and define the external momenta as%
\begin{equation}
p=\mathcal{Q}\frac{\bar{n}}{2}+\frac{m_{N}^{2}}{\mathcal{Q}}\frac{n}%
{2},~\ \ p^{\prime}=\mathcal{Q}\frac{n}{2}+\frac{m_{N}^{2}}{\mathcal{Q}}%
\frac{\bar{n}}{2},~\ \ \mathcal{Q=}Q\frac{1}{2}\left[  1+\sqrt{1+\frac
{4m_{N}^{2}}{Q^{2}}}\right]  =Q+\mathcal{O}(m_{N}^{2}/Q^{2}),
\end{equation}%
\begin{equation}
~\ 2(pp^{\prime})=\mathcal{Q}^{2}+\frac{m_{N}^{4}}{\mathcal{Q}^{2}}\approx
Q^{2}, 
\end{equation}
where $m_{N}$ is the nucleon mass. 
For the incoming and outgoing collinear quarks we always imply%
\begin{equation}
p_{i}=x_{i}\mathcal{Q}\frac{\bar{n}}{2}+p_{\bot i}+\left(  x_{i}^{\prime
}~\frac{m_{N}^{2}}{\mathcal{Q}}\right)  \frac{n}{2},~\ \ ~p_{i}^{\prime}%
=y_{i}\mathcal{Q}\frac{n}{2}+p_{\bot i}^{\prime}+\left(  y_{i}^{\prime}%
~\frac{m_{N}^{2}}{\mathcal{Q}}\right)  \frac{\bar{n}}{2}%
,\ \ \ \label{quark mom}%
\end{equation}
with the transverse momenta%
\be
p_{\bot}^{2}\sim p_{\bot}^{\prime2}\sim\Lambda^{2},
\ee
and where $x_i$ and $x^\prime_i$ denote fractions of the corresponding momentum-component.  
Computing  the Feynman diagrams in Sec.III  we neglect by power suppressed components  and assume
\begin{equation}
p\simeq Q\frac{\bar{n}}{2},~p_{i}\simeq x_{i}p,~\ p^{\prime}\simeq Q\frac
{n}{2},\ ~p_{i}^{\prime}\simeq y_{i}p^{\prime},\ \ \
\end{equation} 
In many formulas  we use convenient notation   $\bar x_{i}=1-x_{i}$. 
We also use the following notation for scalar products%
\begin{equation}
(a\cdot n)\equiv a_{+}~,~(a\cdot\bar{n})\equiv a_{-}\ .
\end{equation}
and Dirac contractions
\begin{equation}
p_{\mu}\gamma^{\mu}\equiv\Dslash{p}\equiv\hat{p}.
\end{equation}

Nucleon FFs are defined as the matrix elements of the e.m. current
between the nucleon states:%
\begin{equation}
\left\langle p^{\prime}\right\vert J_{e.m.}^{\mu}(0)\left\vert p\right\rangle
=\bar{N}(p^{\prime})\left[  \gamma^{\mu}(F_{1}+F_{2})-\frac{(p+p^{\prime
})^{\mu}}{2m_{N}}~F_{2}\right]  N(p),\label{FF:def}%
\end{equation}
with nucleon spinors normalized  as $\bar N N=2m_{N}$.

\section*{Appendix B: Soft correlation function in perturbation theory}

The leading order perturbative expression for the $\bs{S}_{V}^{\text{ud}}$ reads:
\begin{align}
\left(  ~\bs{S}_{V}^{\text{ud}}\right)  _{\text{LO}}  &
=\frac{3}{16\pi^{8}}\int dk_{1}~dk_{2}~\delta\left(
k_{1}^{+}-\omega_{1}\right)  \delta\left(  k_{2}^{+}-\omega_{2}\right)
\delta\left(  k_{1}^{-}-\nu_{1}\right)  \delta\left(  k_{2}^{-}-\nu
_{2}\right)  \\
&
~\ \ \ \ \ \ \ \ \ \ \ \ \ \ \ \ \ \ \ \ \ \ \ \ \ \ \ \ \ \ \ \ \ \ \ \ \ \ \ \frac
{\frac{1}{8}\text{Tr}\left[  \left( \hat k_{1}+m\right)  \gamma_{-}C~\left(
C\gamma_{+}\left(\hat  k_{2}+m\right)  \right)  ^{\text{T}}\right]  }{[\omega
_{1}\nu_{1}-k_{1\bot}^{2}-m^{2}+i\varepsilon][\omega_{2}\nu_{2}-k_{2\bot}%
^{2}-m^{2}+i\varepsilon]}.
\end{align}
The factor $1/8$ in front of trace is chosen for convenience. Calculation of the trace in the  numerator yields:
\be
\frac{1}{8}\text{Tr}\left[  \left(\hat  k_{1}+m\right)  \gamma_{-}C~\left(
C\gamma_{+}\left(\hat  k_{2}+m\right)  \right)  ^{\text{T}}\right]  =-m^{2}+\left(  k_{1\perp}\cdot k_{2\perp
}\right)
\ee
This allows us to write:
\be
\left(  ~\boldsymbol{S}_{V}^{\text{ud}}\right)  _{\text{LO}} 
=-\frac{3m^{2}}{16\pi^{8}}  \frac{1}{4}\int dk_{1\perp}~dk_{2\perp}~
\frac{1}{[\omega_{1}\nu_{1}-k_{1\bot}^{2}-m^{2}+i\varepsilon][\omega_{2}\nu_{2}-k_{2\bot}^{2}-m^{2}+i\varepsilon]}.
\label{SV:1}
\ee
In order to proceed further we must take into account the specific properties of the jet functions. We always assume that the soft fractions $\omega_{i}$ and $\nu_{i}$ are positive.  Mathematically it comes out from the analytical properties of the diagrams and imposes specific restrictions on the integrand in Eq.(\ref{SV:1}).  
In order to see this assume that  $-\infty<\omega_{i}<\infty$ but then we keep the Feynman $i\varepsilon$-prescription in the jet functions.  From calculations of the diagrams in Fig.\ref{jet-f-diagrams} 
one can easily obtain that all denominators of the jet functions in Eqs. (\ref{Jua})-(\ref{Jdb}) are defined with $-i\varepsilon$, for instance
\be
\frac{1}{\omega_{1}+\omega_{2}}\frac1{\omega_{1}}\frac1{\omega_{2}}\rightarrow 
\frac{1}{[\omega_{1}+\omega_{2}-i\varepsilon]}\frac1{[\omega_{1}-i\varepsilon]}\frac1{[\omega_{2}-i\varepsilon]}.
\ee
The same arguments also true for $\nu_{i}$.  Consider now the convolution integrals
\be
\int d\omega_{1}d\omega_{2}\frac{1}{[\omega_{1}+\omega_{2}-i\varepsilon]}\frac1{[\omega_{1}-i\varepsilon]}\frac1{[\omega_{2}-i\varepsilon]} S_{V}(\omega_{i}),
\ee
where   $S_{V}(\omega_{i})$ is represented by expression (\ref{SV:1}). Computing  $d\omega_{i}$  by residues  we obtain that nontrivial results 
originates only from the poles of the propagators in  $S_{V}(\omega_{i})$ in  (\ref{SV:1}). Therefore this allows to us to represent the propagators in  (\ref{SV:1}) 
as   $\delta$-functions:
\begin{align}
\frac{1}{[\omega_{1}\nu_{1}-k_{1\bot}^{2}-m^{2}+i\varepsilon][\omega_{2}%
\nu_{2}-k_{2\bot}^{2}-m^{2}+i\varepsilon]}  =&  (2\pi i)^{2}  \theta(\omega_{i}>0)\theta
(\nu_{i}>0)\\
&\times \delta\left(  \omega_{1}\nu_{1}-k_{1\bot}^{2}-m^{2}\right)  \delta\left(
\omega_{2}\nu_{2}-k_{2\bot}^{2}-m^{2}\right)  .
\end{align} 
Then we obtain%
\begin{align}
\left(  ~\boldsymbol{S}_{V}^{\text{ud}}\right)  _{\text{LO}} &  =
\frac{3m^{2}}{16\pi^{6}} 
\theta(\nu_{i}>0)\int
dk_{1\perp}~dk_{2\perp}~\delta\left(  \omega_{1}\nu_{1}-k_{1\bot}^{2}%
-m^{2}\right)  \delta\left(  \omega_{2}\nu_{2}-k_{2\bot}^{2}-m^{2}\right)  \\
&  =
\frac{3m^{2}}{16\pi^{6}} 
~\theta
(\nu_{i}>0)\theta(\omega_{i}>0)~~\theta(\omega_{1}\eta_{1}>m^{2})\theta
(\omega_{2}\eta_{2}>m^{2}).
\end{align}

\section*{Appendix C: Derivation of the collinear contribution $D^{\mu_{\bot}}_{cp'}$}

Consider first  denominator (\ref{D}). In the collinear region (\ref{collpp}) we obtain:
\begin{align}
&  [\left(  p-k_{1}-k_{2}\right)  ^{2}-m^{2}]
\left(  p-p_{3}-k_{1}-k_{2}\right)^{2}\left(  p-p_{3}-k_{1}\right)  ^{2}
[\left(  p_{1}-k_{1}\right)  ^{2}-m^{2}]
\\
&  \simeq \left[  -Q\left(  k_{1}^{-}+k_{2}^{-}\right)  \right]  \left[
-Q\bar{x}_{3}\left(  k_{1}^{-}+k_{2}^{-}\right)  \right]  \left[  -Q\bar
{x}_{3}k_{1}^{-}\right]  \left[  -Qx_{1}k_{1}^{-}\right] \\
&  \simeq Q^{4}(\bar{x}_{3}^{2}x_{1})\left[  -\left(  k_{1}^{-}+k_{2}%
^{-}\right)  \right]  ^{2}\left[  -k_{1}^{-}\right]  ^{2}.
\end{align}
The remaining propagators are soft,  of order $\Lambda^{2}$. Hence for the denominator we obtain
\be
\textrm{Den}\sim Q^{8}\Lambda^{12}
\label{DenD}
\ee

In the numerator we have
\bea
\mathrm{Num}&=&
\bar \xi_{1}^{\prime}(\hat k_{1}+m)
 \left[\gamma^{j}~\xi_{1} \right]
~\bar{\xi}_{2}^{\prime}~\gamma^{i}(\hat p^{\prime}-\hat p_{3}^{\prime}-\hat k_{1}+m)\gamma^{\alpha}(\hat k_{2}+m) 
 \left[  \gamma^{\beta}(\hat p-\hat p_{3}-\hat k_{1})\gamma^{j}~\xi_{2}\right]
 \\ &&
\bar{\xi}_{3}^{\prime}~\gamma^{\alpha}\left(\hat  p^{\prime}-\hat k_{1}-\hat k_{2}+m\right)
\left[  ~\gamma_{\perp}^{\mu}\left(  \hat p-\hat k_{1}-\hat k_{2}+m\right)  \gamma^{\beta}~\xi_{3}\right] 
  \eea
where we single out by brackets [...] the numerators of the hard propagators and hard gluon vertices. Using that $\hat p \xi \simeq 0$ 
we can rewrite this piece as:
\bea
&& \left[\gamma^{j}\xi_{1} \right]\otimes \left[  \gamma^{\beta}(\hat p-\hat p_{3}-\hat k_{1})\gamma^{j}\xi_{2}\right]\otimes
  \left[  \gamma_{\perp}^{\mu}\left(  \hat p-\hat k_{1}-\hat k_{2}+m\right)  \gamma^{\beta}\xi_{3}\right]
  \nonumber \\
&& \phantom{ \left[\gamma^{j}\xi_{1} \right]\otimes \left[  \gamma^{\beta}(\hat p-\hat p_{3}-k_{1})\gamma^{j}\xi_{2}\right]} 
\simeq (-4)(pk_{1})\left[\gamma^{j}\xi_{1} \right]\otimes \left[  \gamma^{j}\xi_{2}\right]\otimes \left[  \gamma_{\perp}^{\mu}\xi_{3}\right] .
\eea
Then we obtain
\bea
\text{Num}&=&2Q(  -k_{1}^{-}) 
\bar{\xi}_{1}^{\prime}\gamma^{i}\left(\hat  k_{1}+m\right)  \left[  \gamma^{j}\xi_{1}\right]  
\bar{\xi}_{2}^{\prime}\gamma^{i}(\hat p^{\prime}-\hat p_{3}^{\prime}-\hat k_{1}+m)\gamma^{\alpha}(\hat k_{2}+m)
\left[ \gamma^{j}~\xi_{2}\right]  
\nonumber \\ &&
\phantom{2Q(  -k_{1}^{-})  }\times \bar{\xi}_{3}^{\prime}~\gamma^{\alpha}\left( \hat  p^{\prime}-\hat k_{1}-\hat k_{2}+m\right) 
\left[\gamma_{\perp}^{\mu}~\xi_{3}\right]  .
\eea
Recall that   $ k_{1}^{-}  \sim Q$, then
for the remaining terms one finds  
\bea
&\bar{\xi}_{1}^{\prime}\gamma^{i}\left(\hat k_{1\perp}\right)  \left[
\gamma^{j}\xi_{1}\right]  \bar{\xi}_{3}^{\prime}~\gamma^{\alpha}\left(
-\hat k_{1\perp}-\hat k_{2\perp}\right)  ~\left[  \gamma_{\perp}^{\mu}~\xi_{3}\right]
\bar{\xi}_{2}^{\prime}~\gamma^{i}(-\hat k_{1\perp})\gamma^{\alpha}(\hat k_{2\bot
})\left[  \gamma^{j}~\xi_{2}\right]  
\nonumber \\   &
\sim\Lambda^{4}\, \bar{\xi}_{1}^{\prime}\Gamma_{1}\xi_{1}\, \bar{\xi}_{2}^{\prime}\Gamma_{2}\xi_{2}\, \bar{\xi}_{3}^{\prime}\Gamma_{3}\xi_{3} ,
\eea
where $\Gamma_{i}$ denote certain Dirac matrices. Hence one obtains:
\be
\text{Num}\sim Q^{2}\Lambda^{4},
\ee
and  for the whole diagram one finds:
\be
D_{cp^{\prime}}\sim\frac{1}{Q^{6}}~
\bar{\xi}_{1}^{\prime}\Gamma_{1}\xi_{1}\bar{\xi}_{2}^{\prime}\Gamma_{2}\xi_{2}\bar{\xi}_{3}^{\prime}\Gamma_{3}\xi_{3}.
\ee

Taking into account notation in Eqs.(\ref{collPin},\ref{collPout}) the collinear  contribution  can be represented as
\bea
D^{\mu_{\bot}}_{cp^{\prime}}\simeq \mathcal{C}\, \bar\chi'_{\bs \beta}\,
\int dk_{1}^{-}dk_{2}^{-}\, \mathcal{V}_{\bs \beta\bs \beta'}(k_{i}^{-})
\frac{
\left[  \gamma^{j}\right]_{ \beta'_{1}\alpha_{1}} 
\left[  \gamma^{j}\right]  _{\beta'_{2}\alpha_{2}}
\left[ \gamma_{\perp}^{\mu}\right]  _{ \beta'_{3}\alpha_{3}} 
}
{ \bar{x}_{3}^{2}x_{1} \left[  -Q\left(  k_{1}^{-}+k_{2}^{-}\right)\right]  ^{2} \left[  -Qk_{1}^{-}\right]  }\, \chi_{\bs \alpha} ,
\label{collD}
\eea
with
\bea
\mathcal{V}(k_{i}^{-})  &  = &
\frac12 \int dk_{1}^{+}dk_{2}^{+}~dk_{1\bot}dk_{2\bot}
\frac{
\left\{ \gamma^{i}\left(  \hat k_{1}+m \right)\right\}  _{\beta_{1}\beta'_{1}}
 }
{
\left[  k_{2}^{2}-m^{2}\right]  \left[  k_{1}^{2}-m^{2}\right]
}
\nonumber \\
&  &
\times \frac{
\left\{  \gamma^{i}
(\hat p^{\prime}-\hat p_{3}^{\prime}-\hat k_{1}+m )  \gamma^{\alpha}(\hat k_{2}+m)\right\}_{\beta_{2}\beta'_{2}}
\left\{ \gamma^{\alpha}
 \left(\hat p^{\prime}-\hat k_{1}-\hat k_{2}+m \right)    \right\}  _{\beta_{3}\beta'_{3}}
 }
{ \left[ \left(  p^{\prime}-k_{1}-k_{2}\right)  ^{2}-m^{2}\right]  
\left[  \left(  p^{\prime}-p_{3}^{\prime} -k_{1}\right)^{2}-m^{2}\right]  
\left(  k_{1}+k_{2}-p^{\prime}+p_{3}^{\prime} \right) ^{2}
\left(  k_{1}-p_{1}^{\prime}\right)  ^{2} }.
\label{CollV}
\eea
Eq.(\ref{CollV}) yields the contribution to the evolution kernel at 2-loop
approximation. The integral with respect to $k_{1,2}^{-}$ in Eq.(\ref{collD}) can be interpreted as a
convolution integral of leading order hard coefficient function with the given
part of evolution kernel.

\section*{Appendix D:  Calculation of $D^{\mu_{\bot}}_{s}$}
The expression for the soft integral $J_{s}$ in Eq.(\ref{Ds:res})  reads
\be
J_{s} =\frac{1}{Q^{6}}\frac1{x_{1}\bar x_{3}^{2}}\frac1{y_{1}\bar y_{3}^{2}} I_{s},
\ee
with
\begin{align}
I_{s}&  =\int dk_{2\perp}dk_{1\perp}~\int dk_{2}^{+}~dk_{1}^{+}~\frac
{1~~}{~\left[  -k_{1}^{+}-\tau_{+}/y_{1}\right]  \left[  -(k_{1}^{+}+k_{2}%
^{+})-\tau_{+}/\bar{y}_{3}\right]  \left[  -(k_{1}^{+}+k_{2}^{+})\right]  }\\
&  \int dk_{2}^{-}dk_{1}^{-}\frac{1}{\left[  -k_{1}^{-}-\tau_{-}/x_{1}\right]
\left[  -(k_{1}^{-}+k_{2}^{-})-\tau_{-}/\bar{x}_{3}\right]  \left[  -(k_{1}%
^{-}+k_{2}^{-})\right]  ~}~\frac{m^{2}}{\left[  k_{2}^{2}-m^{2}\right]
\left[  k_{1}^{2}-m^{2}\right]  }%
\end{align}
For simplicity let us  redefine the notations as  %
\be
k_{i}^{+}\rightarrow\beta_{i},~k_{i}^{-}\rightarrow\alpha_{i},~dk_{i\perp
}\rightarrow dk_{i},~\ \tau_{1}\equiv\tau_{-}/x_{1},~\ \tau_{3}\equiv\tau_{-}/\bar
{x}_{3},~\tau_{1}^{\prime}\equiv\tau_{+}/y_{1},~\tau_{3}^{\prime}\equiv\tau_{+}
/\bar{y}_{3}.
\ee
So that
\begin{align*}
I_{s}  &  =\int dk_{2}dk_{1}~\int d\beta_{1}~d\beta_{2}~\frac{1~~}{~\left[
-\beta_{1}-\tau_{1}\right]  \left[  -(\beta_{1}+\beta_{2})-\tau_{3}\right]
\left[  -(\beta_{1}+\beta_{2})\right]  }\\
&\times  \int d\alpha_{1}d\alpha_{2}~\frac{1}{\left[  -\alpha_{1}-\tau_{1}^{\prime
}\right]  \left[  -(\alpha_{1}+\alpha_{2})-\tau_{3}^{\prime}\right]  \left[
-(\alpha_{1}+\alpha_{2})\right]  ~}~\frac{m^{2}}
{\left[  \alpha_{2}\beta_{2}-k_{2}^{2}-m^{2}\right]  \left[  \alpha_{1}\beta_{1}-k_{1}^{2}%
-m^{2}\right]  }.
\end{align*}
Expressions in square brackets implies $[...]\equiv [...+i\varepsilon]$. Next we use the same trick as in Appendix A: we integrate over $d\alpha_{1,2}$ by residues, 
rewrite the poles as $\delta$-functions:
\be
\frac1
{\left[  \alpha_{2}\beta_{2}-k_{2}^{2}-m^{2}\right]  \left[  \alpha_{1}\beta_{1}-k_{1}^{2}-m^{2}\right]  }\rightarrow 
(2\pi i)^{2}\theta(\beta_{i})\theta(\alpha_{i})\delta(\alpha_{1}\beta_{1}-k_{1}^{2}-m^{2})\delta( \alpha_{2}\beta_{2}-k_{2}^{2}-m^{2})
\ee
and  integrate over transverse momenta. This yields
\begin{align*}
I_{s}  &  =(2\pi i)^{2} m^{2}\int_{0}^{\infty}d\beta_{1}~d\beta_{2}~\frac{1~~}{~\left[
\beta_{1}+\tau_{1}\right]  \left[  (\beta_{1}+\beta_{2})+\tau_{3}\right]
(\beta_{1}+\beta_{2})}\\
&\times  \int_{0}^{\infty}d\alpha_{1}d\alpha_{2}~\frac{\theta(\alpha_{2}\beta
_{2}>m^{2})\theta(\alpha_{1}\beta_{1}>m^{2})}{\left[  \alpha_{1}+\tau
_{1}^{\prime}\right]  \left[  (\alpha_{1}+\alpha_{2})+\tau_{3}^{\prime
}\right]  (\alpha_{1}+\alpha_{2})~}%
\end{align*}%
 After simple substitutions this integral can be written as
\begin{align*}
I_{s}& =(2\pi i)^{2} m^{2}~\int_{0}^{\infty}d\beta_{2}~\frac{1}{(1+\beta_{2})}\int_{0}%
^{\infty}d\alpha_{1}d\alpha_{2}~\frac{\theta(\alpha_{2}\beta_{2}>m^{2}%
)~\theta(\alpha_{1}>m^{2})}{(\alpha_{1}+\alpha_{2})~}\\
&  ~\ \ \ \ \ \ \ \ \ \ \ \ \ \ \ \times\int_{0}^{\infty}d\beta_{1}\frac
{\beta_{1}~~}{~\left(  \beta_{1}+\tau_{1}\right)  \left[  \beta_{1}%
(1+\beta_{2})+\tau_{3}\right]  }\frac{1}{\left(  \alpha_{1}+\beta_{1}\tau
_{1}^{\prime}\right)  }\frac{1}{\left[  \alpha_{1}+\alpha_{2}+\beta_{1}%
\tau_{3}^{\prime}\right]  }.
\end{align*}
Using that
\be
\frac{\beta_{1}~~}{~\left[  \beta_{1}(1+\beta_{2})+\tau_{3}\right]  }=\frac
{1}{(1+\beta_{2})}-\frac{\tau_{3}}{(1+\beta_{2})\left[  \beta_{1}(1+\beta
_{2})+\tau_{3}\right]  }\simeq\frac{1}{(1+\beta_{2})},
\ee
\begin{align}
\frac{1}{\left(  \alpha_{1}+\beta_{1}\tau_{1}^{\prime}\right)  }\frac
{1}{\left[  \alpha_{1}+\alpha_{2}+\beta_{1}\tau_{3}^{\prime}\right]  }  &
=\frac{1}{\left(  \alpha_{1}+\alpha_{2}\right)  }\left[  \frac{1}{\left(
\alpha_{1}+\beta_{1}\tau_{1}^{\prime}\right)  }-\frac{\tau_{3}^{\prime}%
}{\left[  \alpha_{1}+\alpha_{2}+\beta_{1}\tau_{3}^{\prime}\right]  }\right] \\
&  \simeq\frac{1}{\left(  \alpha_{1}+\alpha_{2}\right)  }\frac{1}{\left(
\alpha_{1}+\beta_{1}\tau_{1}^{\prime}\right)  },
\end{align}
where we neglected small contributions proportional to  infinitesimal mass. Therefore we obtain
\bea
I_{s}&=&(2\pi i)^{2}m^{2}~\int_{0}^{\infty}d\beta_{2}~\frac{1}{(1+\beta_{2})^{2}}\int%
_{0}^{\infty}d\alpha_{1}d\alpha_{2}~\frac{~\theta(\alpha_{1}>m^{2})}{(\alpha_{1}+\alpha_{2})^{2}~}\int%
_{0}^{\infty}d\beta_{1}\frac{\theta(\alpha_{2}\beta_{2}>m^{2})}{~\left(  \beta_{1}+\tau_{1}^{\prime}\tau
_{1}\right)  }\frac{1}{\left(  \alpha_{1}+\beta_{1}\right)  }
\\
&  =&(2\pi i)^{2}m^{2}\int_{m^{2}}^{\infty}~d\alpha_{1}~\frac{\ln\left[  \alpha_{1}%
/\tau_{1}^{\prime}\tau_{1}\right]  }{\alpha_{1}-\tau_{1}^{\prime}\tau_{1}%
}~\int_{0}^{\infty}d\alpha_{2}\frac{1}{(\alpha_{1}+\alpha_{2})^{2}%
(1+m^{2}/\alpha_{2})}\\
&\simeq&
(2\pi i)^{2}\int_{1}^{\infty}~\frac{d\alpha_{1}%
}{\alpha_{1}}\left(  ~\ln\alpha_{1}+\ln\left[  m^{2}/\tau_{1}^{\prime}\tau
_{1}\right]  \right)  \frac{\alpha_{1}-\ln\alpha_{1}-1}{(1-\alpha_{1})^{2}}
\\
&=& (2\pi i)^{2} \ln\left[  \tau_{1}^{\prime}\tau_{1}/m^{2}\right]  \left(  1-\frac{\pi^{2}%
}{6}\right)  + \mathcal{O}(1).
\eea
Therefore we finally arrive at:
\be
I_{s}=(2\pi i)^{2} \ln\left[  \tau_{+}\tau_{-}/m^{2}\right] \left(1-\frac{\pi^{2}}{6}\right)  +\mathcal{O}(1).
\ee

\section*{Appendix E: Calculation of $\Psi^{\bs \beta}_{20}(k_{i}^{-})$}

 Let us rewrite the Eq.(\ref{Psi20}) as 
\be
{ \Psi}^{\bs \beta}_{20}(k_{i}^{-})   = 
\bar \chi'_{ \beta'_1\beta'_2\beta_3}[ \gamma^{i} ]_{\beta'_{1}\beta_{1}}
[ \gamma^{i} ]_{\beta'_{2}\beta_{2}} J_{20},
\label{Psi20A}
\ee
with
\bea
J_{20}&=&
\int dk_{1}^{+}dk_{2}^{+}~dk_{1\bot }dk_{2\bot}
\frac{ m^{2}  }
{  \left[  k_{2}^{2}-m^{2}\right]  \left[  k_{1}^{2}-m^{2}\right]  }
\nonumber \\ &  &
\times\frac{ -2(p' k_{1})}
{ \left[ \left(  p^{\prime}-k_{1}-k_{2}\right)  ^{2}-m^{2}\right]  
\left[  \left(  p^{\prime}-p_{3}^{\prime} -k_{1}\right)^{2}-m^{2}\right]  
\left(  k_{1}+k_{2}-p^{\prime}+p_{3}^{\prime} \right) ^{2}
\left(  k_{1}-p_{1}^{\prime}\right)^{2} }.
\label{J20}
\eea
The exact answer for $J_{20}$ is very complicated. 
We shall compute this expression only in the region of small $k^{-}_{i}\rightarrow 0$
assuming that their ratio is fixed $k^{-}_{1}/k^{-}_{2}\sim \mathcal{O}(1)$. Again, redefine for simplicity 
light-cone decomposition:
\be
k_{i}=k^{-}_{i}\frac{n}{2}+k^{+}_{i}\frac{\bar{n}}{2}+k_{i\bot} \equiv \alpha_{i}\frac{n}{2}+\beta_{i}\frac{\bar{n}}{2}+k_{i},
\ee
and rewrite Eq.(\ref{J20}) as 
\begin{align}
J_{20}  &  =\int d\beta_{1}~d\beta_{2}~dk_{2}dk_{1}
\frac{m^{2}}{\left[  \alpha_{2}\beta_{2}-k_{2}^{2}-m^{2}\right]  \left[
\alpha_{1}\beta_{1}-k_{1}^{2}-m^{2}\right]  }
\nonumber \\ & 
\times \frac{(-Q\beta_{1})}{\left[  (\alpha_{1}-\bar{y}_{3}Q)\beta_{1}-k_{1}
^{2}\right]  \left[  (\alpha_{1}-y_{1}Q)\beta_{1}-k_{1}^{2}\right]  }
\nonumber \\
&\times  \frac{1}{\left[  \left(  \beta_{1}+\beta_{2}\right)  (\alpha_{1}+\alpha
_{2}-Q)-(k_{1}+k_{2})^{2}\right]  \left[  \left(  \beta_{1}+\beta_{2}\right)
(\alpha_{1}+\alpha_{2}-\bar{y}_{3}Q)-(k_{1}+k_{2})^{2}\right]  }.
\label{J20:1}
\end{align}
Performing integrations over $\beta_{1}$ and $\beta_{2}$ by residues we can neglect by all poles
for which
\be
\alpha_{i}>Q-\alpha_{j}\text{ or}~\alpha_{i}>Qy_{i}-\alpha_{j},
\ee
because in this case $\alpha_{i}$  can not be small. Then we have contribution only
from the poles associated with the first two propagators in Eq.(\ref{J20:1}) (in the first line).  
The result can be written as
\begin{align}
J_{20}    &  \simeq (2\pi i)^{2}\theta(0<\alpha_{1}<Q\alpha_{1}^{\max
})~\theta(0<\alpha_{2}<Q\alpha_{2}^{\max})
\nonumber  \\
&\times  \int d\beta_{1}d\beta_{2}dk_{2}dk_{1}\delta(\alpha_{2}\beta_{2}
-k_{2}^{2}-m^{2})\delta(\alpha_{1}\beta_{1}-k_{1}^{2}-m^{2})
\frac{(-Q\beta_{1})~m^{2}}{\mathrm{Den}  }.
\label{J20:2}
\end{align}
with the denominator
\bea
\mathrm{Den}=& {\left[  (\alpha_{1}-\bar{y}_{3}Q)\beta_{1}%
-k_{1}^{2}\right]  \left[  (\alpha_{1}-y_{1}Q)\beta_{1}-k_{1}^{2}\right]  }\left[  \left(  \beta_{1}+\beta_{2}\right)  (\alpha_{1}+\alpha_{2}-Q)-(k_{1}+k_{2})^{2}\right]
\nonumber \\
&  \phantom {\left[  \left(  \beta_{1}+\beta_{2}\right)  (\alpha_{1}+\alpha_{2}-Q)-(k_{1}+k_{2})^{2}\right]  }
\times \left[  \left(  \beta_{1}+\beta_{2}\right)(\alpha_{1}+\alpha_{2}-\bar{y}_{3}Q)-(k_{1}+k_{2})^{2}\right] 
\label{Den}
\eea
The maximal values  $\alpha_{1}^{\max}\sim\alpha_{2}^{\max}\sim1$. Their explicit
values are not important.  The integrand can be further simplified. We can neglect the small fractions $\alpha_{i}\ll Q$
in the denominator (\ref{Den}) that yields
\begin{align}
\mathrm{Den}\simeq & \left[  (-\bar{y}_{3}Q)\beta_{1}-k_{1}^{2}\right]  \left[  (-y_{1}Q)\beta_{1}-k_{1}^{2}\right]  
\nonumber \\ & 
\phantom{ \left[  (-\bar{y}_{3}Q)\beta_{1}-k_{1}^{2}\right]  } 
\times \left[\left(  \beta_{1}+\beta_{2}\right)  (-Q)-(k_{1}+k_{2})^{2}\right]  
\left[\left(  \beta_{1}+\beta_{2}\right)  (-\bar{y}_{3}Q)-(k_{1}+k_{2})^{2}\right] . 
\label{Den:1}
\end{align}
Then we take into account that the dominant contribution arises from the region where
\be
k_{i}^{2}=\alpha_{i}\beta_{i}-m^{2} \ll Q,
\ee
as it follows from the $\delta$-functions in  Eq.(\ref{J20:2}).
Therefore we can  also neglect  the transverse momenta $k_{i}$ in the propagators in Eq.(\ref{Den:1}):
\be
\mathrm{Den}\simeq {\left[  (-\bar{y}_{3}Q)\beta_{1}\right]
\left[  (-y_{1}Q)\beta_{1}\right]  }\frac{1}{\left[  \left(  \beta_{1}%
+\beta_{2}\right)  (-Q)\right]  \left[  \left(  \beta_{1}+\beta_{2}\right)
(-\bar{y}_{3}Q)\right]  }=\frac{(-m^{2})}{y_{1}\bar{y}_{3}^{2}Q^{3}}\frac
{1}{\beta_{1}\left(  \beta_{1}+\beta_{2}\right)  ^{2}}.
\ee
Finally  we obtain
\bea
J_{20}& \simeq&(2\pi i)^{2} \theta(0<\alpha_{1}<Q\alpha_{1}^{\max}
)~\theta(0<\alpha_{2}<Q\alpha_{2}^{\max})\frac{(-1)}{y_{1}\bar{y}_{3}^{2}
Q^{3}}
\nonumber \\ &&
 \times \int d\beta_{1}~d\beta_{2}~dk_{2}dk_{1}~\delta(\alpha_{2}\beta
_{2}-k_{2}^{2}-m^{2})~\delta(\alpha_{1}\beta_{1}-k_{1}^{2}-m^{2})\frac{m^{2}
}{\beta_{1}\left(  \beta_{1}+\beta_{2}\right)  ^{2}}.
\eea
A simple calculation yields:
\bea
J_{20}&\simeq &(2\pi i)^{2} \theta(0<\alpha_{1}<Q\alpha_{1}^{\max}
)~\theta(0<\alpha_{2}<Q\alpha_{2}^{\max})\frac{(-1)}{y_{1}\bar{y}_{3}^{2}
Q^{3}}~\alpha_{2}\ln(1+\alpha_{1}/\alpha_{2})
\nonumber \\
&\equiv &(2\pi i)^{2} \theta(0<k^{-}_{1}/Q<k^{-}_{1max})~\theta(0<k^{-}_{2}/Q<k^{-}_{2max})\frac{(-1)}{y_{1}\bar{y}_{3}^{2}
Q^{3}}~k^{-}_{2}\ln(1+k^{-}_{1}/k^{-}_{2}).
\eea
Substituting this into Eq.(\ref{Psi20A})  we obtain the required result.

\end{document}